\documentclass[american,onecolumn]{revtex4}
\usepackage{amstext}
\usepackage{amssymb}
\usepackage{graphicx}

\makeatletter

\newcommand{\noun}[1]{\textsc{#1}}
\DeclareRobustCommand{\greektext}{%
  \fontencoding{LGR}\selectfont\def\encodingdefault{LGR}}
\DeclareRobustCommand{\textgreek}[1]{\leavevmode{\greektext #1}}
\DeclareFontEncoding{LGR}{}{}
\DeclareTextSymbol{\~}{LGR}{126}
\newcommand{\lyxmathsym}[1]{\ifmmode\begingroup\def\b@ld{bold}
  \text{\ifx\math@version\b@ld\bfseries\fi#1}\endgroup\else#1\fi}

\providecommand{\tabularnewline}{\\}

\@ifundefined{textcolor}{}
{%
 \definecolor{BLACK}{gray}{0}
 \definecolor{WHITE}{gray}{1}
 \definecolor{RED}{rgb}{1,0,0}
 \definecolor{GREEN}{rgb}{0,1,0}
 \definecolor{BLUE}{rgb}{0,0,1}
 \definecolor{CYAN}{cmyk}{1,0,0,0}
 \definecolor{MAGENTA}{cmyk}{0,1,0,0}
 \definecolor{YELLOW}{cmyk}{0,0,1,0}
}

\newtheorem{theorem}{Theorem}
\newtheorem{definition}{Definition}

\makeatother

\usepackage{babel}
\begin{document}

\title{On the origin of inflation by using exotic smoothness }

\author{Torsten Asselmeyer-Maluga}

\email{torsten.asselmeyer-maluga@dlr.de}

\address{German Aero space Center (DLR), Rutherfordstr. 2, 12489 Berlin, Germany}

\author{Jerzy Kr\'ol}

\email{iriking@wp.pl}

\address{University of Silesia, Institute of Physics, ul. Uniwesytecka 4,
40-007 Katowice, Poland}
\begin{abstract}
In this paper we discuss a spacetime having the topology of $S^{3}\times\mathbb{R}$
but with a different smoothness structure leading to a geometric model
for inflation, called geometric inflation. In particular this spacetime
is not globally hyperbolic and we obtain a time line with a spatial
topology change from the 3-sphere to a homology 3-sphere and back.
The topology of the spacetime remains invariant. Among the infinite
possible smoothness structures of this spacetime, we choose a homology
3-sphere constructed from the knot $8_{10}$with hyperbolic geometry,
i.e. admitting a homogenous metric of negative scalar curvature. We
discuss the accelerated expansion for FLRW cosmology caused by the
topology change. In contrast to other inflation models, this process
stops after a finite time. Alternatively, the topology change can
be also described by a $SU(2)-$valued scalar field. Then we calculate
the expansion rate (having more than 60 e-folds) and the energy /
time scale. The coupling to matter is also interpreted geometrically
and the reheating process (as well the supercooled expansion during
inflation) is naturally obtained. The model depends only on a single
parameter, a topological invariant of the homology 3-sphere, and assumes
a Planck size universe of $S^{3}-$topology. The dependence of the
model on the initial state and the a geometric interpretation of quantum
fluctuations are also discussed.
\end{abstract}
\maketitle
\tableofcontents{}

\section{Introduction}

Because of the influx of observational data, recent years have witnessed
enormous advances in our understanding of the early universe. To interpret
the present data, it is sufficient to work in a regime in which spacetime
can be taken to be a smooth continuum as in general relativity, setting
aside fundamental questions involving the deep Planck regime. However,
for a complete conceptual understanding as well as interpretation
of the future, more refined data, these long-standing issues will
have to be faced squarely. As an example, can one show from first
principles that the smooth spacetime of general relativity is valid
at the onset of inflation? In this paper we will focus mainly on this
question about the origin of inflation. Inflation is today the main
theoretical framework that describes the early Universe and that can
account for the present observational data \cite{WMAP-7-years}. In
thirty years of existence \cite{Guth1981,Linde1982}, inflation has
survived, in contrast with earlier competitors, the tremendous improvement
of cosmological data. In particular, the fluctuations of the Cosmic
Microwave Background (CMB) had not yet been measured when inflation
was invented, whereas they give us today a remarkable picture of the
cosmological perturbations in the early Universe. In nearly all known
models, the inflation period is caused by one or more scalar field(s)\cite{InflationBook}.

\subsection{The model}

In this paper we try to derive an inflationary phase from first principles.
The spacetime has the topology of $S^{3}\times\mathbb{R}$ and is
smoothable (i.e. a smooth 4-manifold) \cite{Qui:82} but (we assume)
it is not diffeomorphic to $S^{3}\times\mathbb{R}$. What does it
mean? Every manifold is defined by a collection of charts, the atlas,
including also the transition functions between the charts. From the
physical point of view, charts are the reference frames. The transition
functions define the structure of the manifold, i.e. transition functions
are homeomorphisms (topological manifold) or diffeomorphisms (smooth
manifold). Two (smooth) atlases are compatible (or equivalent) if
their union is a (smooth) atlas again. The equivalence class (the
maximal atlas) is called a differential structure%
\footnote{A smooth atlas defines a smoothness structure.%
}. In dimension smaller than $4$, there is only one differential structure,
i.e. the topology of these manifolds define uniquely its smoothness
properties. In contrast, beginning with dimension $4$ there is the
possibility of more than one differential structure. But 4-manifolds
are really special: here there are many examples of 4-manifolds with
infinite many differential structures (countable for compact and uncountable
for non-compact 4-manifolds including $\mathbb{R}^{4}$). Among these
differential structures there is one exceptional, the standard differential
structure. We will illustrate these standard structure for our spacetime
$S^{3}\times\mathbb{R}$. The 3-sphere $S^{3}$ has an unique differential
structure (the standard differential structure) which extends to $S^{3}\times\mathbb{R}$.
All other differential structures (also called misleadingly ''exotic
smoothness structures'') can never split smoothly into $S^{3}\times\mathbb{R}$.
We denote it by $S^{3}\times_{\theta}\mathbb{R}$. Our main hypothesis
is now:\\
\textbf{Main hypothesis}: \emph{The spacetime has the topology $S^{3}\times\mathbb{R}$
but having the differential structure $S^{3}\times_{\theta}\mathbb{R}$.}\\
In \cite{Fre:79} the first $S^{3}\times_{\theta}\mathbb{R}$ was
constructed and we will use this construction here. The details of
the construction including the foliation can be found in section \ref{sec:exotic-S3xR}.
One starts with a homology 3-sphere $\Sigma$, i.e. a compact 3-manifold
$\Sigma$ with the same homology as the 3-sphere but non-trivial fundamental
group, see for instance \cite{Bre:93}. The Poincare sphere is one
example of a homology 3-sphere. Now we consider the 4-manifold $\Sigma\times[0,1]$
with fundamental group $\pi_{1}(\Sigma\times[0,1])=\pi_{1}(\Sigma)$.
By a special procedure (the plus construction, see \cite{Mil:61,Ros:94}),
one can ''kill'' the fundamental group $\pi_{1}(\Sigma)$ in the interior
of $\Sigma\times[0,1]$. This procedure transforms a non-contractable
closed curve (as element of $\pi_{1}(\Sigma)$) to a contractable
curve (denoted as ''killing the fundamental group'' above). It will
result in a 4-manifold $W$ with boundary $\partial W=-\Sigma\sqcup S^{3}$
($-\Sigma$ is $\Sigma$ with opposite orientation), a so-called cobordism
between $\Sigma$ and $S^{3}$. The gluing $-W\cup_{\Sigma}W$ along
$\Sigma$ with the boundary $\partial(-W\cup_{\Sigma}W)=-S^{3}\sqcup S^{3}$
defines one piece of the exotic $S^{3}\times_{\theta}\mathbb{R}$.
The whole construction can be extended to both directions to get the
desired exotic $S^{3}\times_{\theta}\mathbb{R}$ (see \cite{Fre:79,Kir:89}
and the subsection \ref{sub:Constructing-the-exotic} for the details).
There is one critical point in the construction: the 4-manifold $W$
is not a smooth manifold. As Freedman \cite{Fre:82} showed, the 4-manifold
$W$ always exists topologically but, by a result of Gompf \cite{Gom:89}
(using Donaldson \cite{Don:83}), not smoothly (i.e. it does not exists
as a smooth 4-manifold). The 4-manifold $-W\cup_{\Sigma}W$ is also
non-smoothable and we will get a smoothness structure only for the
whole non-compact $S^{3}\times_{\theta}\mathbb{R}$ (see \cite{Qui:82}).
But $S^{3}\times_{\theta}\mathbb{R}$ contains $-W\cup_{\Sigma}W$
with the smooth cross section $\Sigma$. From the physical point of
view we interpret $-W\cup_{\Sigma}W$ as a time line of a cosmos starting
as 3-sphere changing to the homology 3-sphere $\Sigma$ and changing
back to the 3-sphere%
\footnote{This 3-sphere is very complicated embedded (a so-called wild embedding).
We will discuss the corresponding implications in subsection \ref{sub:Dependence-on-initial-condition-quantum-fluctuation}.%
}. But this process is part of every exotic smoothness structure $S^{3}\times_{\theta}\mathbb{R}$,
i.e. we obtain the mathematical fact\\
\textbf{Fact}\emph{: In the spacetime $S^{3}\times_{\theta}\mathbb{R}$
we have a change of the spatial topology from the 3-sphere to some
homology 3-sphere $\Sigma$ and back but without changing the topology
of the spacetime.}\\
Now we have to discuss the choice of the homology 3-sphere $\Sigma$.
At first, usually every homology 3-sphere is the boundary of a topological,
contractable 4-manifold \cite{Fre:82} but this homology 3-sphere
$\Sigma$ never bounds a \textbf{smooth}, contractable 4-manifold.
Secondly, every homology 3-sphere can be constructed by using a knot
\cite{Rol:76}. One starts with the complement $S^{3}\setminus\left(D^{2}\times K\right)$
of a knot $K$ (a smooth embedding $S^{1}\to S^{3}$) and glue in
a solid torus $D^{2}\times S^{1}$ using a special map (a $\pm1$
Dehn twist). The resulting 3-manifold $\Sigma(K)$ is a homology 3-sphere.
For instance the trefoil knot $3_{1}$ (in Rolfsen notation \cite{Rol:76})
generates the Poincare sphere by this method (with $-1$ Dehn twist). 

Our model\emph{ }$S^{3}\times_{\theta}\mathbb{R}$ starts with a 3-sphere
as spatial topology. Now we will be using a powerful tool for the
following argumentation, Thurstons geometrization conjecture \cite{Thu:97}
proved by Perelman \cite{Per:02,Per:03.1,Per:03.2}. According to
this theory, only the 3-sphere and the Poincare sphere carry a homogenous
metric of constant positive scalar curvature (spherical geometry or
Bianchi IX model) among all homology 3-spheres. Also other homology
3-spheres%
\footnote{The homology 3-sphere must be irreducible and therefore generated
by an irreducible knot, i.e. a knot not splittable in a sum of two
non-trivial knots.%
} are able to admit a homogenous metrics. There is a close relation
between Thurstons geometrization theory and Bianchi models in cosmology
\cite{Anderson,Andersson}. Most of the (irreducible) homology 3-spheres
have a hyperbolic geometry (Bianchi V model), i.e. a homogenous metric
of negative curvature. Here we will choose such a hyperbolic homology
3-sphere. As an example we choose the knot $8_{10}$ leading to the
hyperbolic homology 3-sphere $\Sigma(8_{10})$. This choice is not
arbitrary. According to the construction above, we need a homology
3-sphere which does not bound a smooth, contractable 4-manifold. The
homology 3-sphere $\Sigma(8_{10})$ fulfills this condition. For more
details we refer to subsection \ref{sub:Constructing-the-exotic}.

\subsection{Results of the Paper}

At first we will give an overview of our assumptions:
\begin{enumerate}
\item The spacetime is $S^{3}\times_{\theta}\mathbb{R}$ (with\emph{ }topology
$S^{3}\times\mathbb{R}$) containing a homology 3-sphere $\Sigma$
(as cross section).
\item This homology 3-sphere $\Sigma$ is a hyperbolic 3-manifold (with
negative scalar curvature).
\end{enumerate}
According to the mathematical facts above, the 3-sphere is changed
to $\Sigma$ and back in $S^{3}\times_{\theta}\mathbb{R}$, i.e. we
obtain a topology change of the spatial cosmos. In the next section
we will discuss the implication for the cosmology (using the Robertson-Walker
metric). Then we obtain the relation
\[
\dot{a}(t)\cdot\left(\frac{\rho_{0}}{a^{2}}+2\ddot{a}\right)>0
\]
for the scaling parameter $a(t)$ leading to an accelerated expansion
$\ddot{a}>0$ for $\frac{\rho_{0}}{a^{2}}\ll1$. After a short excursion
in the differential topology of exotic 4-manifold, we will construct
explicitly the foliation of $S^{3}\times_{\theta}\mathbb{R}$ and
discuss the physical properties like causality and global hyperbolicity
in section \ref{sec:exotic-S3xR}. The main results can be found in
section \ref{sec:properties-of-inflation}. At first we will show
that the change of the 3-sphere $S^{3}$ to $\Sigma$ can be described
by a $SU(2)-$valued scalar field with double well potential as interaction.
Then we will calculate the expansion rate by analyzing the topology-changing
process. This process is determined by one parameter
\[
\vartheta=\frac{3\cdot vol(\Sigma)}{2\cdot CS(\Sigma)}
\]
which only depend on 3-manifold $\Sigma$. The parameter $\vartheta$
is a relation between two topological invariants of $\Sigma$, the
Chern-Simons invariant $CS(\Sigma)$ (see the appendix \ref{sec:Chern-Simons-invariant})
and the volume $vol(\Sigma)$ (an invariant for a hyperbolic 3-manifold
using Mostow rigidity \cite{Mos:68}, subsection \ref{sub:Mostow-rigidity}).
Then we obtain the expansion rate
\[
l_{sclae}=\exp\left(\vartheta\right)
\]
the energy scale 
\[
e_{scale}=1+\frac{\vartheta}{2}+\frac{\vartheta^{2}}{4}+\frac{\vartheta^{3}}{3!}
\]
the time scale 
\[
t_{scale}=\sum_{n=0}^{5}\frac{\vartheta^{n}}{n!}
\]
and the decrease of the temperature ($T_{0}$ start temperature and
$T_{1}$ end temperature)
\[
\frac{T_{0}}{T_{1}}=e_{scale}
\]
during the inflation (supercooled expansion). For the concrete model
of a hyperbolic 3-manifold $\Sigma=\Sigma(8_{10})$, generated by
the knot $8_{10}$, we obtain the following values:\\

\begin{tabular}{|c|c|c|c|c|}
\hline 
$\vartheta$ & $l_{scale}$ & $e_{scale}$ & $t_{scale}$ & $\frac{T_{0}}{T_{1}}$\tabularnewline
\hline 
\hline 
$83.131...$. & $1.3\cdot10^{36}$ & $115172.2606\ldots$ & $3.5..\cdot10^{7}$ & $115172.2606\ldots$\tabularnewline
\hline 
\end{tabular}\\
\\
If we assume a Planck state (a state of Planck size having Planck
energy along Planck time) then we will obtain the measurable values\\

\begin{tabular}{|c|c|c|c|}
\hline 
size after inflation & Energy scale & time scale & supercooled temperature $(T_{0}=)$\tabularnewline
\hline 
\hline 
$\approx10\, m$ & $\approx10^{14}\, GeV$ & $\approx10^{-37}s$ & $T_{1}\approx10^{27}K$\tabularnewline
\hline 
\end{tabular}\\
\\
Finally we will also discuss the coupling to the matter and the reheating
after inflation. Furthermore, we will obtain the model of parametric
resonance naturally.

\section{FLRW cosmology for a topology change\label{sec:FLR-cosmology}}

The analysis of the WMAP data do not exclude the case that our universe
is a compact 3-manifold with a slightly positive curvature \cite{WMAPcompactSpace2008}.
So, we are able to choose the topology of the spacetime to be $S^{3}\times\mathbb{R}$.
But we remark that $\mathbb{R}^{4}$ topology of the spacetime can
be also chosen. Clearly this spacetime admits a Lorentz metric (given
by the topological condition to admit a non-vanishing vector field
normal to $S^{3}$). But we weaken the condition of global hyperbolicity
otherwise it induces a diffeomorphism\cite{BernalSanchez2003,BernalSanchez2006}
to $S^{3}\times\mathbb{R}$. We will discuss the implications in subsection
\ref{sub:Global-hyperbolicity-and-handles}. The spacetime has the
topology of $S^{3}\times\mathbb{R}$ and is smoothable (i.e. a smooth
4-manifold) \cite{Qui:82} but (we assume) it is not diffeomorphic
to $S^{3}\times\mathbb{R}$. This exotic $S^{3}\times\mathbb{R}$
will be denoted by $S^{3}\times_{\theta}\mathbb{R}$. As stated above
and explained in section \ref{sec:exotic-S3xR}, this $S^{3}\times_{\theta}\mathbb{R}$
contains a homology 3-sphere and we have a topology change from $S^{3}$
to $\Sigma$ and back. Now we will study the geometry and topology
changing process more carefully. Let us consider the Robertson-Walker
metric (with $c=1$)
\[
ds^{2}=dt^{2}-a(t)^{2}h_{ik}dx^{i}dx^{k}
\]
in Friedmann\textendash{}Lema�tre\textendash{}Robertson\textendash{}Walker
(FLRW) cosmology with the scaling function $a(t)$. This metric do
not depend on the topology of the spacetime. One assumes only the
local splitting $N\times[0,1]$ with $N$ a 3-manifold with homogenous
metric and coordinates $x^{i}$ and a (finite) time variable $t$.

At first we assume a spacetime $S^{3}\times\mathbb{R}$ with increasing
function $a(t)$ fulfilling the Friedman equations
\begin{eqnarray}
\left(\frac{\dot{a}(t)}{a(t)}\right)^{2}+\frac{k}{a(t)^{2}} & = & \kappa\frac{\rho}{3}\label{eq:friedman-1}\\
2\left(\frac{\ddot{a}(t)}{a(t)}\right)+\left(\frac{\dot{a}(t)}{a(t)}\right)^{2}+\frac{k}{a(t)^{2}} & = & -\kappa p\label{eq:friedman-2}
\end{eqnarray}
derived from Einsteins equation
\begin{equation}
R_{\mu\nu}-\frac{1}{2}g_{\mu\nu}R=\kappa T_{\mu\nu}\label{eq:Einstein-equation}
\end{equation}
with the gravitational constant $\kappa$ and the energy-momentum
tensor of a perfect fluid
\begin{equation}
T_{\mu\nu}=(\rho+p)u_{\mu}u_{\nu}-pg_{\mu\nu}\label{eq:perfect-fluid-EM-tensor}
\end{equation}
with the (time-dependent) energy density $\rho$ and the (time-dependent)
pressure $p$. The spatial cosmos has the scalar curvature $^{3}R$
\begin{equation}
^{3}R=\frac{3k}{a^{2}}\label{eq:spatial-curvature}
\end{equation}
from the 3-metric $h_{ik}$ and we obtain the 4-dimensional scalar
curvature $R$
\begin{equation}
R=\frac{6}{a^{2}}\left(\ddot{a}\cdot a+\dot{a}^{2}+k\right)\:.\label{eq:4-dim-scalar-curvature}
\end{equation}
Let us consider the model $S^{3}\times\mathbb{R}$ with positive spatial
curvature $k=+1$. In case of dust matter ($p=0$) only, one obtains
a closed universe. Now we consider our model of an exotic $S^{3}\times_{\theta}\mathbb{R}$.
As explained above, the foliation of $S^{3}\times_{\theta}\mathbb{R}$
must contain a hyperbolic homology 3-sphere $\Sigma(8_{10})$ (with
negative scalar curvature). But then we have a transition from a space
with positive curvature to a space with negative curvature and back.
To model this behavior, we consider a time-dependent parameter $k(t)$
in the curvature 
\[
^{3}R(t)=\frac{3k(t)}{a^{2}}
\]
with the following conditions:
\begin{enumerate}
\item The change of the geometry from spherical $k>0$ to hyperbolic $k<0$
happens at $t_{0}$,
\item $k(t)>0$ for $t\ll t_{0}$ and $t\gg t_{0}$.
\end{enumerate}
The change of the topology is an abrupt process. So at first we will
take the path $\Sigma(8_{10})\times(t_{0},t_{0}+\Delta t)$ for a
suitable time interval and discuss the equation (\ref{eq:friedman-1})
for $t>t_{0}$:
\[
-1=k(t_{0}+\Delta t)=\frac{\rho_{0}}{a(t_{0}+\Delta t)}-\left(\dot{a}(t_{0}+\Delta t)\right)^{2}
\]
and the path $S^{3}\times(t_{0}-\Delta t,t_{0})$ for $t<t_{0}$:
\[
+1=k(t_{0}-\Delta t)=\frac{\rho_{0}}{a(t_{0}-\Delta t)}-\left(\dot{a}(t_{0}-\Delta t)\right)^{2}
\]
The difference of these equations (representing the equation at $t_{0}$)
is given by
\begin{eqnarray*}
0 & > & \lim_{\Delta t\to0}\frac{k(t_{0}+\Delta t)-k(t_{0}-\Delta t)}{2\Delta t}=\frac{d}{dt}\left(\frac{\rho_{0}}{a(t)}-\left(\dot{a}(t)\right)^{2}\right)|_{t=t_{0}}\\
 &  & =-\dot{a}(t)\cdot\left(\frac{\rho_{0}}{a^{2}}+2\ddot{a}\right)\:.
\end{eqnarray*}
Now lets take $0\approx\frac{\rho_{0}}{2a^{2}}\ll1$ with good accuracy
and $\dot{a}>0$ (expansion) then we obtain an accelerated expansion
$\ddot{a}>0$. The calculation above is true for every kind of function
$k(t)$, jumping from $+1$ for $t<t_{0}$ to $-1$ for $t>t_{0}$.
This model is only a motivation for a model with a spatial topology
change. For instance, the energy-momentum tensor has to change also
and we will analyze it in subsection \ref{sub:Matter-coupling}.

Now we will reverse the argumentation. Let us assume a solution with
accelerated expansion. If we interpret the equation (\ref{eq:friedman-1})
as a relation to calculate $k(t)$ then the corresponding function
$k(t)$ has to reflect the properties discussed above. In particular,
we will consider a model with a balance between dust matter $(p=0)$
(having the scaling behavior $\rho\sim a^{-3}$) and the expansion.
For $t\ll t_{0}$ and $t\gg t_{0}$ we have a cosmos of constant radius.
As we argue above, in the period around $t_{0}$ we obtained an accelerated
expansion. 
\begin{figure}
\includegraphics[scale=0.4]{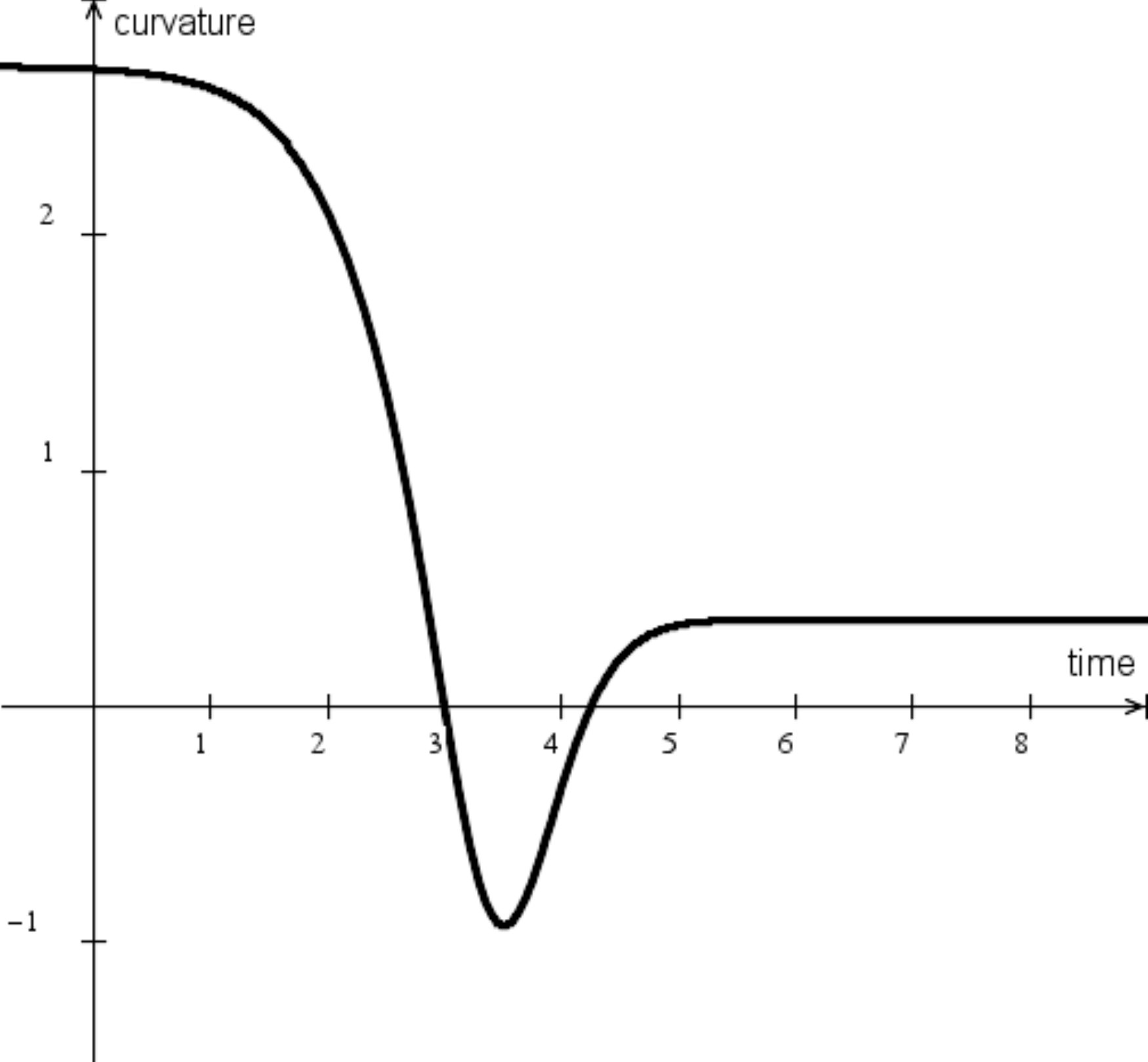} \qquad\includegraphics[scale=0.4]{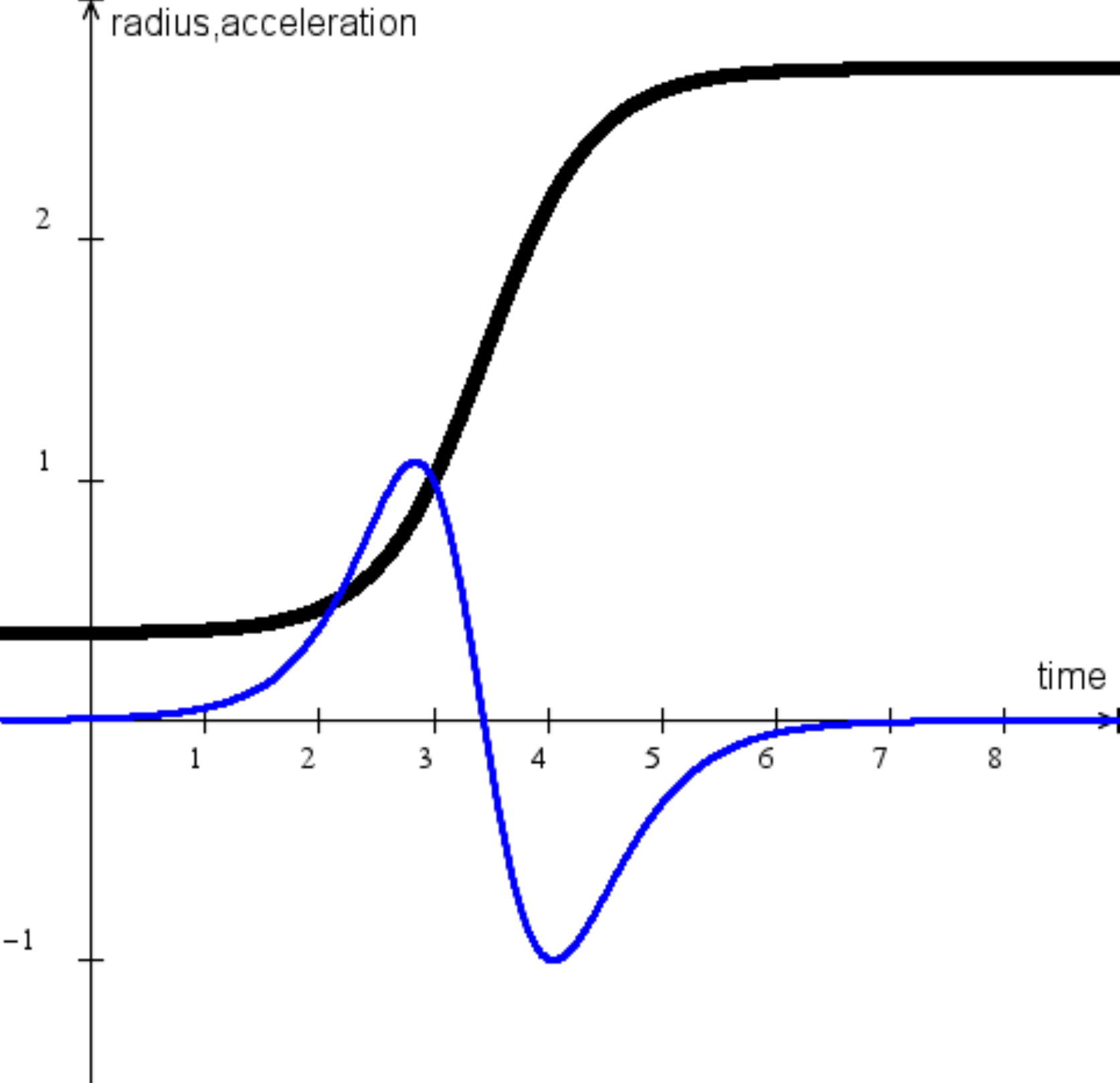}

\caption{time-dependent curvature $k(t)$ with $t_{0}=3,\zeta=1,\rho_{0}=1$(left
fig.) and the solution $a(t)$ (thick curve) as well the acceleration
$\ddot{a}(t)$ (right fig.)\label{fig:time-dependent-curvature-solution}}
\end{figure}
 Now we choose a solution of the Friedmann equation (\ref{eq:friedman-1})
with an accelerated expansion but with a (nearly) constant radius
for $t\ll t_{0}$ and $t\gg t_{0}$. To reflect all these properties
in one function, we choose the solution
\begin{equation}
a(t)=\exp\left(\tanh(\zeta(t-t_{0}))\right)\label{eq:solution}
\end{equation}
also visualized in the right figure of Fig. \ref{fig:time-dependent-curvature-solution}.
Now we reverse the argumentation and interpret the Friedmann equation
to calculate the curvature function $k(t)$ 
\[
k(t)=\frac{\rho_{0}}{a(t)}-\left(\dot{a}(t)\right)^{2}
\]
for dust matter. Of course, this approach is only a approximation
to visualize our model. Finally we obtain

\begin{eqnarray*}
k(t) & = & \rho_{0}\exp\left(-\tanh(\zeta(t-t_{0}))\right)-\\
 &  & -\zeta^{2}\left(1-\tanh^{2}(\zeta(t-t_{0}))\right)^{2}\exp\left(2\cdot\tanh(\zeta(t-t_{0}))\right)\:.
\end{eqnarray*}
This function is plotted in the left figure of Fig. \ref{fig:time-dependent-curvature-solution}
for special parameters confirming the conditions above and the topology
change (encoded into the change of the curvature). In particular we
obtain again \emph{a positive acceleration} $\ddot{a}(t)$ in some
time interval. Furthermore the acceleration $\ddot{a}(t)$ is also
negative or the \emph{inflation process stops}. Finally the topology
change of the space
\begin{equation}
\mbox{spherical 3-sphere}\longrightarrow\mbox{hyperbolic homology 3-sphere}\label{eq:change}
\end{equation}
produces an accelerated expansion of $a(t)$. We will later prove
that the expansion rate has to be exponential which motivates to denote
this model as \emph{inflation.} But in contrast to the usual inflation
models, we derive this behavior from first principles using the spacetime
$S^{3}\times_{\theta}\mathbb{R}$ (with a non-standard differential
structure). Furthermore \emph{in our inflation model, the growing
stops, i.e. our inflation is not eternal}. In subsection \ref{sub:Curvature-contribution-of-CH},
we will discuss the reason for this behavior.

\section{Preliminaries: The exotic $S^{3}\times\mathbb{R}$\label{sec:exotic-S3xR}}

In this section we will describe the construction and some properties
of the exotic $S^{3}\times_{\theta}\mathbb{R}$. Some background can
be found in the book \cite{Asselmeyer2007} (suitable for physicists)
or in the math books \cite{Kir:89,GomSti:1999}.

\subsection{Smoothness on manifolds}

If two manifolds are homeomorphic but non-diffeomorphic, they are
\textbf{exotic} to each other. The smoothness structure is called
an \textbf{exotic smoothness structure}.

The implications for physics are tremendous because we rely on the
smooth calculus to formulate field theories. Thus different smoothness
structures have to represent different physical situations leading
to different measurable results. But it should be stressed that \emph{exotic
smoothness is not exotic physics.} Exotic smoothness is a mathematical
structure which should be further explored to understand its physical
relevance.

Usually one starts with a topological manifold $M$ and introduces
structures on them. Then one has the following ladder of possible
structures:
\begin{eqnarray*}
\mbox{Topology}\to & \mbox{\mbox{piecewise-linear(PL)}}\to & \mbox{Smoothness}\to\\
\qquad\to & \mbox{bundles, Lorentz, Spin etc.}\to & \mbox{metric, geometry,...}
\end{eqnarray*}
We do not want to discuss the first transition, i.e. the existence
of a triangulation on a topological manifold.  The following basic
facts should the reader keep in mind for any $n-$dimensional manifold
$M^{n}$:
\begin{enumerate}
\item The maximal differentiable atlas $\mathcal{A}$ of $M^{n}$ is the
smoothness structure.
\item To determine a smoothness structure it suffices to give a single maximal
differentiable atlas. Thus $\mathbb{R}^{n}$ has an unique smoothness
structure containing the identity map of $\mathbb{R}^{n}$ (\emph{standard
smoothness structure}, $(\mathbb{R}^{n},id_{\mathbb{R}^{n}})$ is
the atlas). 
\item It is difficult to define the \emph{standard smoothness structure}
on a general 4-manifold $M$. One way to get around this difficulty
is the usage of the instability of all exotic smoothness structures
in dimension 4. Stable smoothness structures are able to extend from
a smoothness structure on $M$ to $M\times\mathbb{R}^{k}$ (see \cite{KirSie:77}
for the notation of a stable CAT structure). The classification theory
of smoothness structure \cite{Mun:60,KirSie:77} for all manifolds
of dimension greater than $5$ implies (together with a result of
Quinn \cite{Qui:82} about the vanishing of $\pi_{4}(TOP/O)=0$) that
the smoothness structure of $M\times\mathbb{R}^{k}$ is unique for
all $k>0$ (up to diffeomorphisms). Here we have to assume that the
Kirby-Siebenmann invariant vanishes. We call this smoothness structure
the standard smoothness structure of $M\times\mathbb{R}^{k}$. Then
one can extend this smoothness structure to $M$ by restriction. All
other possible smoothness structures non-diffeomorphic to the standard
smoothness structure are called \emph{exotic} smoothness structures.
\item The existence of a smoothness structure is \emph{necessary} to introduce
Riemannian or Lorentzian structures on $M$, but the smoothness structure
do not further restrict the Lorentz structure.
\end{enumerate}
We want to close this subsection with a general remark: the number
of non-diffeomorphic smoothness structures is finite for all dimensions
$n\not=4$ \cite{KirSie:77}. In dimension four there are many examples
of compact 4-manifolds with infinite finite and many examples of non-compact
4-manifolds with uncountable infinite many non-diffeomorphic smoothness
structures.

\subsection{Global hyperbolicity and decomposition by handles\label{sub:Global-hyperbolicity-and-handles}}

Before we start the investigation of the proposed model, we will discuss
some more general physical implications. Firstly we consider the existence
of a Lorentz metric, i.e. a 4-manifold $M$ (the spacetime) admits
a Lorentz metric if (and only if) there is a non-vanishing vector
field. In case of a compact 4-manifold $M$ we can use the Poincare-Hopf
theorem to state: a compact 4-manifold admits a Lorentz metric if
the Euler characteristic vanishes $\chi(M)=0$. But in a compact 4-manifold
there are closed time-like curves (CTC) contradicting the causality
or more exactly: the chronology violating set of a compact 4-manifold
is non-empty (Proposition 6.4.2 in \cite{HawEll:94}). Non-compact
4-manifolds $M$ admit always a Lorentz metric and a special class
of these 4-manifolds have an empty chronology violating set. If $\mathcal{S}$
is an acausal hypersurface in $M$ (i.e., a topological hypersurface
of $M$ such that no pair of points of $M$ can be connected by means
of a causal curve), then $D^{+}(\mathcal{S})$ is the future Cauchy
development (or domain of dependence) of $\mathcal{S}$, i.e. the
set of all points $p$ of $M$ such that any past-inextensible causal
curve through $p$ intersects $\mathcal{S}$. Similarly $D^{-}(\mathcal{S})$
is the past Cauchy development of $\mathcal{S}$. If there are no
closed causal curves, then $\mathcal{S}$ is a Cauchy surface if $D^{+}(\mathcal{S})\cup\mathcal{S}\cup D^{-}(\mathcal{S})=M$.
As shown in \cite{BernalSanchez2003}, the existence of a Cauchy surface
implies that $M$ is diffeomorphic to $\mathcal{S}\times\mathbb{R}$
. 

This strong result is also connected with the concept of global hyperbolicity.
A spacetime manifold $M$ without boundary is said to be \emph{globally
hyperbolic} if the following two conditions hold:
\begin{enumerate}
\item \emph{Absence of naked singularities}: For every pair of points $p$
and $q$ in $M$, the space of all points that can be both reached
from $p$ along a past-oriented causal curve and reached from $q$
along a future-oriented causal curve is compact.
\item \emph{Chronology}: No closed causal curves exist (or \emph{Causality}
holds on $M$).
\end{enumerate}
Usually condition 2 above is replaced by the more technical condition
\emph{Strong causality holds on $M$} but as
shown in \cite{BernalSanchez2007} instead of \emph{strong
causality} one can write simply the condition \emph{causality}
(and strong causality will hold under causality plus condition 1 above). 

Then together with the diffeomorphism between $M$ and $\mathcal{S}\times\mathbb{R}$
we can conclude that all (non-compact) 4-manifolds $\mathcal{S}\times\mathbb{R}$
are the only 4-manifolds which admit a globally hyperbolic Lorentz
metric \cite{BernalSanchez2003}. The existence of a Cauchy surface
$\mathcal{S}$ implies global hyperbolicity of the spacetime and its
unique representation by $\mathcal{S}\times\mathbb{R}$ (up to diffeomorphism).
But as shown in \cite{BernalSanchez2005}, also the metric is determined
(up to isometry) by global hyperbolicity.

\begin{theorem}\label{thm:global-hyp}

If a spacetime $(M,g)$ is globally hyperbolic, then it is isometric
to $(\mathbb{R}\times\mathcal{S},- f\cdot d\tau^{2}+g_{\tau})$
with a smooth positive function $f:\mathbb{R}\to\mathbb{R}$ and a
smooth family of Riemannian metrics $g_{\tau}$ on $\mathcal{S}$
varying with $\tau$. Moreover, each $\left\{ t\right\} \times\mathcal{S}$
is a Cauchy slice.

\end{theorem} Furthermore in \cite{BernalSanchez2006} it was shown:
\begin{itemize}
\item If a compact spacelike submanifold with boundary of a globally hyperbolic
spacetime is acausal then it can be extended to a full Cauchy spacelike
hypersurface $\mathcal{S}$ of $M$, and
\item for any Cauchy spacelike hypersurface $\mathcal{S}$ there exists
a function as in Th. \ref{thm:global-hyp} such that $\mathcal{S}$
is one of the levels $\tau=constant$.
\end{itemize}
But what are the implications of global hyperbolicity in the exotic
case? At first, the existence of a Lorentz metric is a purely topological
condition which will be fulfilled by all non-compact 4-manifolds independent
of the smoothness structure. But by considering global hyperbolicity
the picture changes. An exotic spacetime $M=\mathcal{S}\times_{\theta}\mathbb{R}$
homeomorphic to $\mathcal{S}\times\mathbb{R}$ is \emph{not diffeomorphic}
to $\mathcal{S}\times\mathbb{R}$. The Cauchy surface $\mathcal{S}$
is a 3-manifold with an unique smoothness structure (up to diffeomorphisms)
-- the standard structure. So, the smooth product $\mathcal{S}\times\mathbb{R}$
must be admit the standard smoothness structure. But the diffeomorphism
\cite{BernalSanchez2003} between $M$ and $\mathcal{S}\times\mathbb{R}$
is necessary for global hyperbolicity. Therefore an \emph{exotic $\mathcal{S}\times_{\theta}\mathbb{R}$
is never globally hyperbolic but admits a Lorentz metric}. Generally
we have an exotic \emph{$\mathcal{S}\times_{\theta}\mathbb{R}$} with
a Lorentz metric such that the projection $\mathcal{S}\times_{\theta}\mathbb{R}\to\mathbb{R}$
is a time-function (that is, a continuous function which is strictly
increasing on future directed causal curves). But then the exotic
\emph{$\mathcal{S}\times_{\theta}\mathbb{R}$} has no closed causal
curves and must contain naked singularities%
\footnote{Any non-compact manifold $M$ admits stably causal metrics (that is,
those with a time function). So, if $M$ is not diffeomorphic to some
product$\mathcal{S}\times\mathbb{R}$, all these (causally well behaved)
metrics must contain naked singularities. We thank M. S\'anchez for
the explanation of this result.%
}. 

\begin{figure}
\includegraphics[angle=90,scale=0.25]{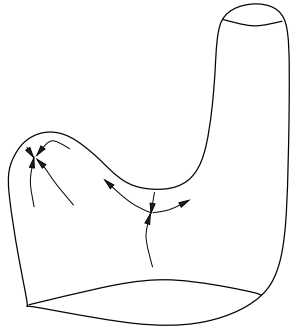}

\caption{Two naked singularities.\foreignlanguage{english}{\label{fig:toy-naked-singularity}}}
\end{figure}

\begin{figure}
\includegraphics{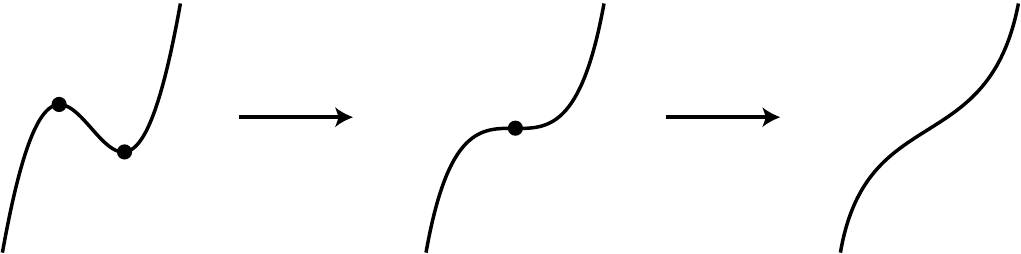}

\caption{Killing a 0- and a 1-handle.\foreignlanguage{english}{\label{fig:killling-handles}}}
\end{figure}

With this result in mind, one should ask for the physical interpretation
of naked singularities. To visualize the problem, we consider the
following toy model: a non-trivial surface (see Fig. \ref{fig:toy-naked-singularity})
connecting two circles which can be deformed to the usual cylinder.
 This example can be described by the concept of a cobordism. A cobordism
$(W,M_{1},M_{2})$ between two $n-$manifolds $M_{1},M_{2}$ is a
$(n+1)-$manifold $W$ with $\partial W=M_{1}\sqcup M_{2}$ (ignoring
the orientation). Then there exists a smooth function $f:W\to[0,1]$
with isolated critical points (vanishing first derivative) such that
$f^{-1}(0)=M_{1},\, f^{-1}(1)=M_{2}$. By general position arguments,
one can assume that all critical points of $f$ occur in the interior
of $W$. In this setting $f$ is called a Morse function on a cobordism.
For every critical point of $f$ (vanishing first derivative) one
adds a so-called $k-$handle $D^{k}\times D^{n-k}$. In our example
in Fig. \ref{fig:toy-naked-singularity}, we add a 2-handle $D^{2}\times D^{0}$
(the maximum) and a 1-handle $D^{1}\times D^{1}$ (the saddle). But
obviously this cobordism is diffeomorphic to the trivial cobordism
$S^{1}\times[0,1]$ (the two boundary components are diffeomorphic
to each other). Therefore the 2-/1-handle pair is ''killed''
in this case. The 2-handle and the 1-handle differ in one direction
whereas the Morse function has a maximum for the 2-handle and a minimum
for the 1-handle. The left graph of Fig. \ref{fig:killling-handles}
visualizes this fact. 
 Furthermore the sequence of graphs from the left to right represents
the process to ''kill'' the handle pair. 

In our example of an exotic $\mathcal{S}\times_{\theta}\mathbb{R}$,
we consider a (non-compact) cobordism between homeomorphic boundary
components (the two Cauchy surfaces at infinity, i.e. ''$\mathcal{S}\times\left\{ -\infty\right\} $
and $\mathcal{S}\times\left\{ +\infty\right\} $''). By a result of
\cite{Mil:65}, a cobordism of this kind (a so-called h-cobordism)
contains only handles of complement dimension in the interior. But
these handles can be killed: the details of the construction can be
found in \cite{Mil:65}. Here we will give only some general remarks.
Any $0-/1-$handle pair as well any $n-/(n+1)-$handle pair (remember
the h-cobordism is $n+1$-dimensional) can be killed by a general
procedure. The killing of a $k-/(k+1)-$handle pair depends on a special
procedure, the Whitney trick. For 4- and 5-dimensional h-cobordisms
(between 3- and 4-manifolds, respectively) we cannot use the Whitney
trick. This failure lies at the heart of the problem to classify 3-
and 4-manifolds (see \cite{GomSti:1999}).

In case of the spacetime we are interested in a 4-dimensional h-cobordism
between 3-manifolds. Here we can kill the $0-/1-$and the $3-/4-$handle
pair of the h-cobordism. As a result we get extra $1-/2-$handle and
$2-/3-$handle pairs. If the Whitney trick works in this case, we
can also kill these pairs of handles. But it is known that the Whitney
trick only works topologically \cite{Fre:82}. The exotic manifolds
$\mathcal{S}\times_{\theta}\mathbb{R}$ (as non-compact examples)
are counterexamples that the (infinite) pairs of handles never cancel
each other. Lets start with $1-/2-$handle pair. The critical point
of the Morse function with index $1$ (the Morse function has a minimum
in one directions (saddle point)) corresponds to the 1-handle. Then
the Morse function of index 2 corresponds to the 2-handle. Each pair
of handles is connected to each other, i.e. the direction representing
the minimum of a 1-handle is connected with one direction representing
the maximum of the 2-handle. This connection between the 1- and the
2-handle is non-trivial, i.e. there are extra intersection points
between the (attaching) sphere of the 1-handle and the (belt) sphere
of the 2-handle (see \cite{GomSti:1999} Prop. 4.2.9). The $2-/3-$handle
pair is dual to the $1-/2-$handle pair, i.e. the number of $2-/3-$handle
pairs is equal to the number of $1-/2-$handle pairs. Each handle
of the pair represents the neighborhood of a naked singularity. Then
these singularities appear only pairwise. The Morse vector field (gradient
of the Morse function) vanishes but we get a saddle having negative
curvature (\emph{no curvature singularity} or \emph{quasi-regular
singularity,} also called \emph{locally extendible singularity} \cite{Ellis1977}).
There is growing evidence for the appearance of this singularity without
violating causality (causal continuity, see \cite{Dowker1997}). Thus
it is very probable that the property of exotic smoothness requiring
the appearance of naked singularities in form of non-canceling 2-handle
pairs can be interpreted in a consistent physically way: In \cite{AsselmeyerRose2012}
we merged exotic smoothness with fermions and bosons obtaining the
consequence that exotic smoothness implies naked singularities as
well as particles. We do not think this is an accident -- we conjecture
that the naked singularities can be seen as the geometrically consequence
of the particles.

\subsection{Constructing the exotic $S^{3}\times\mathbb{R}$\label{sub:Constructing-the-exotic}}

Let $\Sigma$ be a homology 3-sphere which do not bound a contractable,
smooth 4-manifold. According to Freedman \cite{Fre:82}, every homology
3-sphere bounds a contractable, topological 4-manifold but not every
of these 4-manifolds is smoothable. Now we consider the following
pieces: $W_{1}$ is a cobordism between $\Sigma$ and its one-point
complement $\Sigma\setminus pt.$ and $W_{2}$ is a cobordism between
$\Sigma\setminus pt.$ and $\Sigma\setminus pt.$. The manifold $W=\ldots\cup-W_{2}\cup-W_{2}\cup-W_{1}\cup W_{1}\cup W_{2}\cup W_{2}\cup\ldots$
(see \cite{Fre:79}) is homeomorphic to $S^{3}\times\mathbb{R}$ (using
the proper h-cobordism theorem in \cite{Fre:82}) but not diffeomorphic
to it, i.e. $W=S^{3}\times_{\theta}\mathbb{R}$. The construction
of the pieces $W_{1},W_{2}$ rely heavily on the concept of a Casson
handle. We do not plan to overload this subsection and shift the definition
of a Casson handle to the \ref{sec:Casson-handle}. 

Now we consider a Casson handle $CH$ and its 3-stage tower $T_{3}^{0}$.
By the embedding theorem of Freedman (see \cite{Fre:79}, Theorem
1), one can construct another 3-stage tower $T_{3}^{1}$ inside of
$T_{3}^{0}$ (increase the number of self-intersections of the core).
This process can be done infinitely. Lets take the example of an homology
3-sphere $\Sigma$ constructed below. The fundamental group is generated
by one generator. The attachment of one 2-handle transforms this generator
(a non-contractable curve) into a contractable curve but produces
a non-trivial 2-handle. This 2-handle can be eliminated by attaching
a 3-handle. At the end we obtain a trivial fundamental group and a
2-/3-handle pair which can be canceled by using a single Casson handle
by defining an embedding $T_{3}^{0}\hookrightarrow\Sigma\times[0,1]$.
Now by removing an arc $T_{3,arc}^{0}=T_{3}^{0}\setminus\left\{ \mbox{arc}\right\} $
and a line $T_{3,line}^{0}=T_{3}^{0}\setminus\left\{ \mbox{line}\right\} $
in each 3-stage tower we can form the desired cobordisms $W_{1},W_{2}$
above: $W_{1}=\Sigma\times[0,1]\setminus\bigcap_{i=0}^{\infty}T_{3,arc}^{i}$
and $W_{2}=\Sigma\times[0,1]\setminus\bigcap_{i=0}^{\infty}T_{3,line}^{i}$
completing the construction of the exotic $S^{3}\times\mathbb{R}$.
Furthermore we obtain the smooth cross section $\Sigma$ in the part
$-W_{1}\cup W_{1}$ of $W$.

\begin{figure}
\includegraphics[scale=0.5]{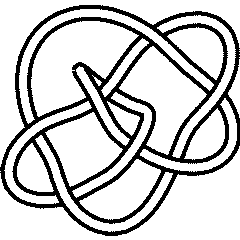}

\caption{Knot $8_{10}$\label{fig:Knot-8_10}}
\end{figure}

For completeness, we will construct one example of a hyperbolic homology
3-sphere which does not bound a contractable 4-manifold. Then we can
use the procedure above to get an exotic $S^{3}\times_{\theta}\mathbb{R}$.
For that purpose we consider a knot $K$, i.e. a smooth embedding
$S^{1}\to S^{3}$. This knot $K$ can be thicken to $N(K)=K\times D^{2}$
and one obtains the knot complement $C(K)=S^{3}\setminus N(K)$ with
boundary $\partial C(K)=T^{2}$. By the attachment of a solid torus
$D^{2}\times S^{1}$ using a $-1$ Dehn twist, one obtains the manifold
$\Sigma=C(K)\cup(D^{2}\times S^{1})$, a homology 3-sphere \cite{Rol:76}.
The geometry of $\Sigma$ is determined by the knot complement $C(K)$.
A fundamental result of Thurston \cite{Thu:97} states that most knot
complements are hyperbolic 3-manifolds. Examples are the figure-8
knot $4_{1}$, the 3-twist knot $5_{2}$ or the knot $8_{10}$ (in
Rolfsen notation \cite{Rol:76}). Therefore we have to look for a
knot inducing a hyperbolic knot complement and leading to a homology
3-sphere which does not bound a smooth contractable 4-manifold. As
Freedman \cite{Fre:82} showed, every homology 3-sphere bound a contractable,
topological 4-manifold. But Donaldson \cite{Don:83} found the first
example, the Poincare sphere, of a homology 3-sphere which fails to
do it. The Poincare sphere is generated by the trefoil knot $3_{1}$
using the procedure above. Every homology 3-sphere homology-cobordant
to the Poincare sphere has the same property \cite{Gordon1975}. From
the knot-theoretical point of view, we have to look for a knot concordant
to the trefoil knot \cite{Livingston1981}. One example is given by
the knot $8_{10}$ (see \cite{Livingston2004} and Fig. \ref{fig:Knot-8_10}).
Then the homology 3-sphere $\Sigma$ constructed from this knot has
all desired properties. 

\subsection{Intermezzo: Geometric structures on 3-manifolds and Mostow rigidity\label{sub:Mostow-rigidity}}

\begin{figure}
\includegraphics{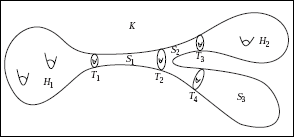}

\caption{Torus (JSJ-) decomposition, $H_{i}$ hyperbolic manifold, $S_{i}$
Graph-manifold, $T_{i}$ Tori \label{fig:Torus-decomposition}}
\end{figure}

A connected 3-manifold $N$ is prime if it cannot be obtained as a
connected sum of two manifolds $N_{1}\#N_{2}$ (see the appendix \ref{sec:Connected-and-boundary-connected}
for the definition) neither of which is the 3-sphere $S^{3}$ (or,
equivalently, neither of which is the homeomorphic to $N$). Examples
are the 3-torus $T^{3}$ and $S^{1}\times S^{2}$ but also the Poincare
sphere. According to \cite{Mil:62}, any compact, oriented 3-manifold
is the connected sum of an unique (up to homeomorphism) collection
of prime 3-manifolds (prime decomposition). A subset of prime manifolds
are the irreducible 3-manifolds. A connected 3-manifold is irreducible
if every differentiable submanifold $S$ homeomorphic to a sphere
$S^{2}$ bounds a subset $D$ (i.e. $\partial D=S$) which is homeomorphic
to the closed ball $D^{3}$. The only prime but reducible 3-manifold
is $S^{1}\times S^{2}$. For the geometric properties (to meet Thurstons
geometrization theorem) we need a finer decomposition induced by incompressible
tori. A properly embedded connected surface $S\subset N$ is called
2-sided%
\footnote{The \textquoteleft{}sides\textquoteright{} of $S$ then correspond
to the components of the complement of $S$ in a tubular neighborhood
$S\times[0,1]\subset N$.%
} if its normal bundle is trivial, and 1-sided if its normal bundle
is nontrivial. A 2-sided connected surface $S$ other than $S^{2}$
or $D^{2}$ is called incompressible if for each disk $D\subset N$
with $D\cap S=\partial D$ there is a disk $D'\subset S$ with $\partial D' =\partial D$.
The boundary of a 3-manifold is an incompressible surface. Most importantly,
the 3-sphere $S^{3}$, $S^{2}\times S^{1}$ and the 3-manifolds $S^{3}/\Gamma$
with $\Gamma\subset SO(4)$ a finite subgroup do not contain incompressible
surfaces. The class of 3-manifolds $S^{3}/\Gamma$ (the spherical
3-manifolds) include cases like the Poincare sphere ($\Gamma=I^{*}$
the binary icosaeder group) or lens spaces ($\Gamma=\mathbb{Z}_{p}$
the cyclic group). Let $K_{i}$ be irreducible 3-manifolds containing
incompressible surfaces then we can $N$ split into pieces (along
embedded $S^{2}$)
\begin{equation}
N=K_{1}\#\cdots\#K_{n_{1}}\#_{n_{2}}S^{1}\times S^{2}\#_{n_{3}}S^{3}/\Gamma\,,\label{eq:prime-decomposition}
\end{equation}
where $\#_{n}$ denotes the $n$-fold connected sum and $\Gamma\subset SO(4)$
is a finite subgroup. The decomposition of $N$ is unique up to the
order of the factors. The irreducible 3-manifolds $K_{1},\ldots,\, K_{n_{1}}$
are able to contain incompressible tori and one can split $K_{i}$
along the tori into simpler pieces $K=H\cup_{T^{2}}G$ \cite{JacSha:79}
(called the JSJ decomposition). The two classes $G$ and $H$ are
the graph manifold $G$ and hyperbolic 3-manifold $H$ (see Fig. \ref{fig:Torus-decomposition}).
The hyperbolic 3-manifold $H$ has a torus boundary $T^{2}=\partial H$,
i.e. $H$ admits a hyperbolic structure in the interior only. One
property of hyperbolic 3- and 4-manifolds is central: \noun{Mostow
rigidity}. As shown by Mostow \cite{Mos:68}, every hyperbolic $n-$manifold
$n>2$ has this property: \emph{Every diffeomorphism (especially every
conformal transformation) of a hyperbolic $n-$manifold is induced
by an isometry.} Therefore one cannot scale a hyperbolic 3-manifold.
Then the volume $vol(\:)$ and the curvature are topological invariants
but for later usages we combine the curvature and the volume into
the Chern-Simons invariant $CS(\:)$ (see appendix \ref{sec:Chern-Simons-invariant}).
Together with the prime and JSJ decomposition
\[
N=\left(H_{1}\cup_{T^{2}}G_{1}\right)\#\cdots\#\left(H_{n_{1}}\cup_{T^{2}}G_{n_{1}}\right)\#_{n_{2}}S^{1}\times S^{2}\#_{n_{3}}S^{3}/\Gamma\,,
\]
we can discuss the geometric properties central to Thurstons geometrization
theorem: \emph{Every oriented closed prime 3-manifold can be cut along
tori (JSJ decomposition), so that the interior of each of the resulting
manifolds has a geometric structure with finite volume.} Now, we have
to clarify the term ''geometric structure''. A model geometry is a
simply connected smooth manifold $X$ together with a transitive action
of a Lie group $G$ on $X$ with compact stabilizers. A geometric
structure on a manifold $N$ is a diffeomorphism from $N$ to $X/\lyxmathsym{\textgreek{G}}$
for some model geometry $X$, where $\lyxmathsym{\textgreek{G}}$
is a discrete subgroup of $G$ acting freely on $X$. t is a surprising
fact that there are also a finite number of three-dimensional model
geometries, i.e. 8 geometries with the following models: spherical
$(S^{3},O_{4}(\mathbb{R}))$, Euclidean $(\mathbb{E}^{3},O_{3}(\mathbb{R})\ltimes\mathbb{R}^{3})$,
hyperbolic $(\mathbb{H}^{3},O_{1,3}(\mathbb{R})^{+})$, mixed spherical-Euclidean
$(S^{2}\times\mathbb{R},O_{3}(\mathbb{R})\times\mathbb{R}\times\mathbb{Z}_{2})$,
mixed hyperbolic-Euclidean $(\mathbb{H}^{2}\times\mathbb{R},O_{1,3}(\mathbb{R})^{+}\times\mathbb{R}\times\mathbb{Z}_{2})$
and 3 exceptional cases called $\tilde{SL}_{2}$ (twisted version
of $\mathbb{H}^{2}\times\mathbb{R}$), NIL (geometry of the Heisenberg
group as twisted version of $\mathbb{E}^{3}$), SOL (split extension
of $\mathbb{R}^{2}$ by $\mathbb{R}$, i.e. the Lie algebra of the
group of isometries of the 2-dimensional Minkowski space). We refer
to \cite{Scott1983} for the details.

\subsection{The foliation of the exotic $S^{3}\times_{\theta}\mathbb{R}$\label{sub:foliation-of-exotic-S3xR}}

A delicate structure for an exotic 4-manifolds is a codimension-1
foliation. In appendix \ref{sec:Foliation-foliated-cobordism}, we
give a short account in foliations and foliated cobordism including
a construction of a codimension-1 foliation of $S^{3}$ using a polygon
in the hyperbolic plane $\mathbb{H}^{2}$. 

\begin{figure}
\includegraphics[scale=0.75]{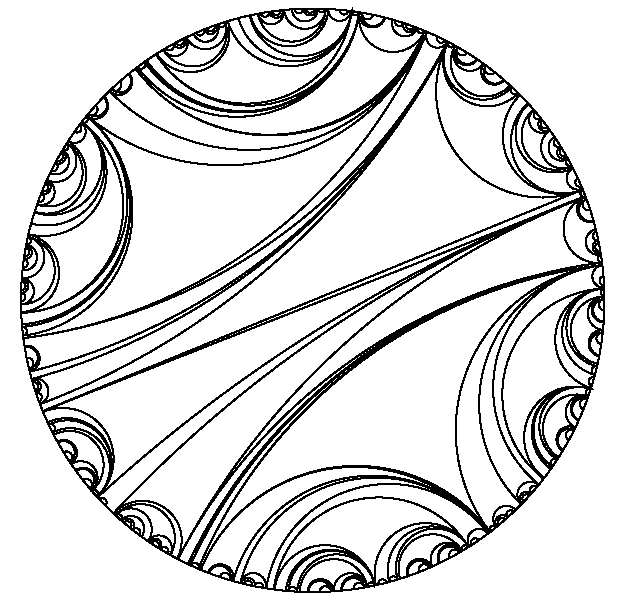}

\caption{(geodesic) lamination of the disk\label{fig:lamination-disk}}
\end{figure}

Let $W=S^{3}\times_{\theta}\mathbb{R}$ be an exotic non-compact 4-manifold
admitting a codimension-1 foliation. But the exoticness of $W$ restricts
the possible foliations. At first we do not have the obvious foliation
with leaves $S^{3}\times\left\{ t\right\} $ and $t\in\mathbb{R}$,
otherwise we can use the unique smoothness structure of the 3-sphere
to induce the standard smoothness structure on $W$ contradicting
the exoticness. Furthermore, by the same reasons, there is no smoothly
embedded 3-sphere in $W$. For the construction of the foliation we
consider the following decomposition of the 3-sphere: 
\[
S^{3}=\left(S^{3}\setminus\bigcup_{n}\left(D^{2}\times S^{1}\right)\right)\cup_{T^{2}}\bigcup_{n}\left(D^{2}\times S^{1}\right)
\]
used in the Dehn surgery (see subsection \ref{sub:Constructing-the-exotic}).
In the notation of the previous subsection, the manifold $W_{1}$
is a cobordism between $\Sigma$ and $\Sigma\setminus pt.$ which
is homeomorphic \cite{Fre:82} to a cobordism between $\Sigma$ and
$S^{3}\setminus pt.$ In the construction of $\Sigma$ above we used
Dehn surgery along the knot $K=8_{10}$. Dually we can decompose $\Sigma$
like
\[
\Sigma=\left(S^{3}\setminus\left(D^{2}\times S^{1}\right)\right)\cup_{T^{2},-1}\left(K\times D^{2}\right)
\]
where $\cup_{T^{2},-1}$ denotes the $-1$ Dehn twist. Now we choose
the $n-$component link 
\[
L_{\Sigma}=K\sqcup_{n-1}S^{1}
\]
and write for $\Sigma$
\[
S^{3}=\left(S^{3}\setminus\bigcup_{n}\left(D^{2}\times S^{1}\right)\right)\cup_{T^{2},-1}\left(D^{2}\times L_{\Sigma}\right)\:.
\]
Therefore $\Sigma$ and $S^{3}$ agreed on the subset $\left(S^{3}\setminus\bigcup_{n}\left(D^{2}\times S^{1}\right)\right)$
and we are able to construct a cobordism $W_{3}$ with
\[
\partial W_{3}=\left(S^{3}\setminus\bigcup_{n}\left(D^{2}\times S^{1}\right)\right)\sqcup-\left(S^{3}\setminus\bigcup_{n}\left(D^{2}\times S^{1}\right)\right)\,.
\]
A foliation of $W_{3}$ cannot be the trivial one, otherwise it will
contradict the exotic smoothness structure. But we can choose a foliation
as a foliated cobordism inducing a foliation of $\left(S^{3}\setminus\bigcup_{n}\left(D^{2}\times S^{1}\right)\right)$.
From the topological point of view, $W_{3}$ is given by
\[
W_{3}=\left(S^{3}\setminus\bigcup_{n}\left(D^{2}\times S^{1}\right)\right)\times[0,1]\:.
\]
The foliation of $W_{3}$ is induced from the foliation of the boundary,
i.e. a foliation of $\left(S^{3}\setminus\bigcup_{n}\left(D^{2}\times S^{1}\right)\right)$.
But this foliation is constructed from a foliation of a polygon $P$
with $n$ vertices in the hyperbolic plane $\mathbb{H}^{2}$ (see
appendix \ref{sec:Foliation-foliated-cobordism}). We will describe
this foliation in the appendix. These foliations are related to (geodesic)
laminations of the disk (see Fig. \ref{fig:lamination-disk} for an
example). 
 The leaves of the foliation of the polygon $P$ are curves $\gamma$
starting and ending at the boundary. Then the leaves of the foliation
of $\left(S^{3}\setminus\bigcup_{n}\left(D^{2}\times S^{1}\right)\right)$
are of the form$\gamma\times S^{1}$. Finally the leaves of $W_{3}$
are $\gamma\times S^{1}\times[0,1]$. The remaining components of
the exotic $W=S^{3}\times_{\theta}\mathbb{R}$ are a cobordism $W(L_{\Sigma})$
between $D^{2}\times L_{\Sigma}$ and $\bigcup_{n}D^{2}\times S^{1}$
equipped with Reeb foliations (see appendix \ref{sec:Foliation-foliated-cobordism}
and Fig. \ref{fig:foliation-torus}). We will later see that this
cobordism is the topological expression for the creation of matter
(see subsection \ref{sub:Matter-coupling}).

\section{The properties of the inflation model and its description\label{sec:properties-of-inflation}}

In the model above, we assume implicitly that the transition from
$k=+1$ to $k=-1$ is an abrupt process. But topology never claimed
how fast is a concrete transition. Therefore we have to analyze the
structure of the spacetime (i.e. $S^{3}\times_{\theta}\mathbb{R}$)
from the geometrical and topological point of view to determine this
transition.

In the following we will use the strategy:
\begin{enumerate}
\item The exotic $S^{3}\times_{\theta}\mathbb{R}$ is build from a hyperbolic
homology 3-sphere $\Sigma$ which does not bound a smooth contractable
4-manifold. Part of the construction is a suitable embedded Casson
handle (or at least a 3-stage tower). We refer to appendix \ref{sec:Casson-handle}
for a description of the Casson handle.
\item The change from $S^{3}$ to $\Sigma$ is described by infinite pairs
of $1-/2-$handles (or dually by $2-/3-$handle pairs) which is canceled
by the Casson handle allowing a Morse-theoretic description by using
a scalar field.
\item Every stage of the Casson handle is a collection of immersed 2-disks.
The immersed disk is also described by a canceling $1-/2-$handle
pair. From the Morse-theoretical point of view, every immersed disk
is a pair of saddle points with a contribution to the negative curvature.
\item For every stage $N$ of the Casson handle we obtain a contribution
\[
\frac{1}{N!}\left(\frac{3\cdot Vol(\Sigma)}{CS(\Sigma)}\right)^{N}
\]
to the curvature, which is expressed by the Chern-Simons invariant
(see \ref{sec:Chern-Simons-invariant}). The combinatorial factor
is related to the embedding of the Casson handle (using a hyperbolic
metric).
\item The embedding of the Casson handle uses a hyperbolic metric and we
obtain an exponential increase of the curvature. The whole process
(starting with the 3-sphere) stops after the formation of $\Sigma$
(by using Mostow rigidity, see subsection \ref{sub:Mostow-rigidity}).
If the 3-sphere has radius $L_{P}$ (the Planck length) the we will
obtain for the radius $a_{0}$ of $\Sigma$
\[
a_{0}=L_{P}\cdot\exp\left(\frac{3\cdot Vol(\Sigma)}{2\cdot CS(\Sigma)}\right)
\]

\end{enumerate}

\subsection{Morse-theoretic description of the topological transition $S^{3}\to\Sigma$\label{sub:morse-theoretic-description}}

As explained above the transition $S^{3}\to\Sigma$ (written as cobordism
$W_{1}$, see above) is an expression of the exotic smoothness structure.
Every topological transition is reflected in the critical values of
the Morse function. The difference between $S^{3}$ and $\Sigma$
is represented by the fundamental group $\pi_{1}(S^{3})=0$ and $\pi_{1}(\Sigma)\not=0$.
Every generator of the fundamental group $\pi_{1}(\Sigma)$ can be
''killed'' by a pair of 2-/3-handles producing two critical values
in the Morse function (of index 2 and 3, respectively). These critical
points are naked singularities (see the discussion in subsection \ref{sub:Global-hyperbolicity-and-handles}).
The Morse vector field (gradient of the Morse function) vanishes but
we get a saddle having negative curvature (\emph{no curvature singularity}
or \emph{quasi-regular singularity,} also called \emph{locally extendible
singularity} \cite{Ellis1977}). There is growing evidence for the
appearance of this singularity without violating causality (causal
continuity, see \cite{Dowker1997}). A pair of 2-/3-handles can be
canceled making the interior of the cobordism $W_{1}$ to a simple-connected
4-manifold. But to start the cancellation process we need a special
embedded disk or better a Casson handle. But the existence of the
exotic smoothness structure forbids the cancellation of the handle
pairs in the Casson handle. The generic case is given by the following
model. Clue a 1-/2-handle pair (dual to a 2-/3-handle pair) to a 0-handle.
The 0-handle is modeled by a minimum in Morse theory. But the 1-/2-handle
pair generates a maximum and a minimum which can be canceled (see
Fig. \ref{fig:killling-handles}). Unfortunately exotic smoothness
forbids this cancellation and we obtain an extra pair of maximum/minimum.
Let us consider the commutative diagram, 
\begin{eqnarray}
\Sigma & \stackrel{\Psi}{\longrightarrow} & \mathbb{R}\nonumber \\
\phi\downarrow & \circlearrowright & \updownarrow id\label{eq:commuting-diagram-1}\\
S^{3} & \stackrel{\psi}{\longrightarrow} & \mathbb{R}\nonumber 
\end{eqnarray}
The map $\psi$ is a Morse function of the 3-sphere $S^{3}$ and $\Psi$
is the Morse function of $\Sigma$. At the 3-sphere there are only
two critical points: one maximum and one minimum (the two poles).
A short calculation of the homology groups of the cobordism $W_{1}$
between $S^{3}$ and $\Sigma$ shows that the homology-groups $H_{0}(M)=H_{3}(M)=\mathbb{Z}$
are the only non-trivial groups of the cobordism. Furthermore the
evaluation of the exact sequence of the pair $(W_{1},\,\partial W_{1})$
gives the result that $H_{0}(W_{1})$ is generated by one of the boundary
components, e.g. by $H_{0}(S^{3})$ and $H_{3}(W_{1})$ by the other
one, i.e. $H_{3}(\Sigma)$. The transition $y=\phi(x)$ represented
by $W_{1}$ maps the Morse function $\psi(y)=||y||^{2}$ on $S^{3}$
to the Morse function $\Psi(x)=||\phi(x)||^{2}$. 

A field-theoretic discussion of the Morse-theoretic result starts
with the following simple idea: The 3-sphere is isomorph to $SU(2)$
and so one can define a formal group-operation on $S^{3}$. Then the
map $\phi:\Sigma\to S^{3}$ is a map $\phi:\Sigma\to SU(2)=\mathbb{R}\otimes SU(2)$
which can be interpreted as a $SU(2)$-valued scalar-field over $\Sigma$.
Witten has derived the Morse-theory from the field-theoretic properties
\cite{Wit:82a}. His ansatz gives a field-theoretic construction (a
$\sigma$ model) of the dynamics leading to a ``tunneling path''
between two critical points. In our case, the field of the Witten-construction
is the $SU(2)$-scalar $\phi$ over $\Sigma$. The 1-/2-handle pair
above and the 0-handle is represented by a Morse function with two
minima and one maxima:
\[
\psi(y)=||y||^{4}-||y||^{2}
\]
on $S^{3}$. But we describe the transition $S^{3}\to\Sigma$ by a
$SU(2)$-valued scalar field $\phi:\Sigma\to SU(2)$ and obtain
\[
\Psi(x)=\phi(x)^{4}-\phi(x)^{2}
\]
the Morse function at $\Sigma$ (where we substitute $||\phi||^{n}$
by $\phi^{n}$). We remark that there is a certain freedom in the
choice of the potential. Here we assume the equality of the two minima
but a potential like
\[
\Psi(x)=\phi(x)^{4}-\phi(x)^{3}-\phi(x)^{2}
\]
is also possible. Currently we have no idea to assign the potential.
But we interpret this Morse function as a potential for the scalar
field and obtain the expression
\begin{equation}
\mathcal{L}_{\phi}=D_{\mu}\phi\cdot D^{\mu}\phi+V(\phi)=D_{\mu}\phi\cdot D^{\mu}\phi+\frac{\rho_{\phi}}{2}(\phi^{4}-\phi^{2})\:,\label{eq:Lagrangian-scalar-field}
\end{equation}
as Lagrangian of the scalar field where $\rho_{\phi}$ is the energy
density of $\phi$ which is directly related to the curvature of the
Morse function. The coupling to gravity can be realized by adding
the Einstein-Hilbert action. Obviously this model has a symmetry-breaking
phase whose meaning will be investigated in a forthcoming paper. If
we are able to cancel the handle pair then the Morse function is reduced
to the Morse function for the 0-handle, i.e.
\[
\psi(y)=||y||^{2}
\]
and we obtain the potential $V(\phi)=\phi^{2}$ in the Lagrangian
(\ref{eq:Lagrangian-scalar-field}). The cancellation process can
be used to obtain the energy of symmetry breaking (see subsection
\ref{sub:Energy-time-scale}).

\subsection{Curvature-contribution of the Casson handle\label{sub:Curvature-contribution-of-CH}}

The idea of a Casson handle can be simply expressed: to kill the $1-/2-$handle
pair one needs an embedded disk. But in dimension 4, it fails and
one obtains an immersed disk (a disk with self-intersections) only.
This immersed disk (also represented by a canceling $1-/2-$handle
pair) can be changed to an embedded disk by embedding a disk again
etc. The Casson handle is the expression of this procedure ad infinitum.
For more details about Casson handles we refer to appendix \ref{sec:Casson-handle}.

The following points are central in the following argumentation: firstly
the appearance of the Chern-Simons invariant defined by the scalar
curvature (therefore related to the geometry) and secondly the usage
of an infinite construction, the so-called Casson handle, to obtain
the topological structure of the transition $S^{3}$ to $\Sigma$.
The first point, the Chern-Simons invariant, will determine the scale
factor and the second point, the Casson handle, will give the exponential
behavior. 

\emph{Step 1}: As shown by Witten \cite{Wit:89.2,Wit:89.3,Wit:91.2},
the action
\begin{equation}
\intop_{\Sigma}\,^{3}R\sqrt{h}\, d^{3}x=L\cdot CS(\Sigma)\label{eq:Witten-relation}
\end{equation}
for every 3-manifold $\Sigma$ is related to the Chern-Simons invariant
$CS(\Sigma)$ (see \ref{sec:Chern-Simons-invariant}). We argued above
that, because of Mostow rigidity, $\Sigma$ has an invariant volume.
Then the scaling factor $L$ is independent of the volume and we obtain
\begin{equation}
L\cdot CS(\Sigma_{0,1},A)=L^{3}\cdot\frac{CS(\Sigma)}{L^{2}}=\intop_{\Sigma}\frac{CS(\Sigma)}{L^{2}\cdot vol(\Sigma)}\,\sqrt{h}\, d^{3}x\label{eq:CS-integral-relation}
\end{equation}
by using
\[
L^{3}\cdot vol(\Sigma)=\intop_{\Sigma}\sqrt{h}\, d^{3}x\,.
\]
Together with (\ref{eq:spatial-curvature}), we can compare the kernels
of the integrals (\ref{eq:Witten-relation}) and (\ref{eq:CS-integral-relation})
to get for every time 
\[
\frac{3k}{a^{2}}=\frac{CS(\Sigma)}{L^{2}\cdot vol(\Sigma)}\,.
\]
Finally we obtain the scaling factor
\begin{equation}
\vartheta=\frac{a^{2}}{L^{2}}=\frac{3\cdot vol(\Sigma)}{|CS(\Sigma)|}\label{eq:scaling-CH}
\end{equation}
where we set $k=-1$ which cancels the negative sign of $CS(\Sigma)$.
We set $CS(\Sigma)$ instead of $|CS(\Sigma)|$ in the following.

\emph{Step 2}: The hyperbolic structure (induced by the hyperbolic
structure of $\Sigma$) is an (geometric) expression for an accelerated
expansion. It is an amazing fact that 4-dimensional hyperbolic structures
show also Mostow rigidity (see subsection \ref{sub:Mostow-rigidity}).
Then we obtain a constant term in the action and in the Friedmann
equation 
\[
\left(\frac{\dot{a}}{a}\right)^{2}=\frac{1}{L^{2}}
\]
with respect to the length scale $L$ of the hyperbolic structure.
In the following we will work with quadratic expressions because we
will mainly argue with the curvatures. Then we obtain
\begin{equation}
da^{2}=\frac{a^{2}}{L^{2}}\, dt^{2}=\vartheta\, dt^{2}\label{eq:scale-quadratic-expansion}
\end{equation}
by using the scale $\vartheta$.

\begin{figure}
\includegraphics[scale=0.25]{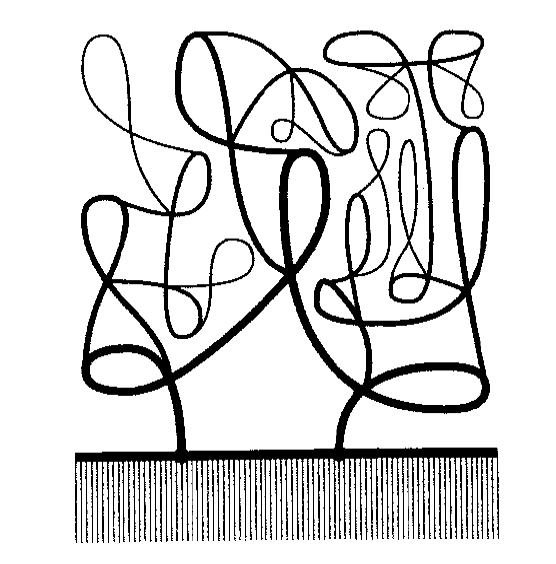}

\caption{schematic picture of the first four stages of a Casson handle\label{fig:schematic-picture-CH}}
\end{figure}

\begin{figure}
\includegraphics[scale=0.1]{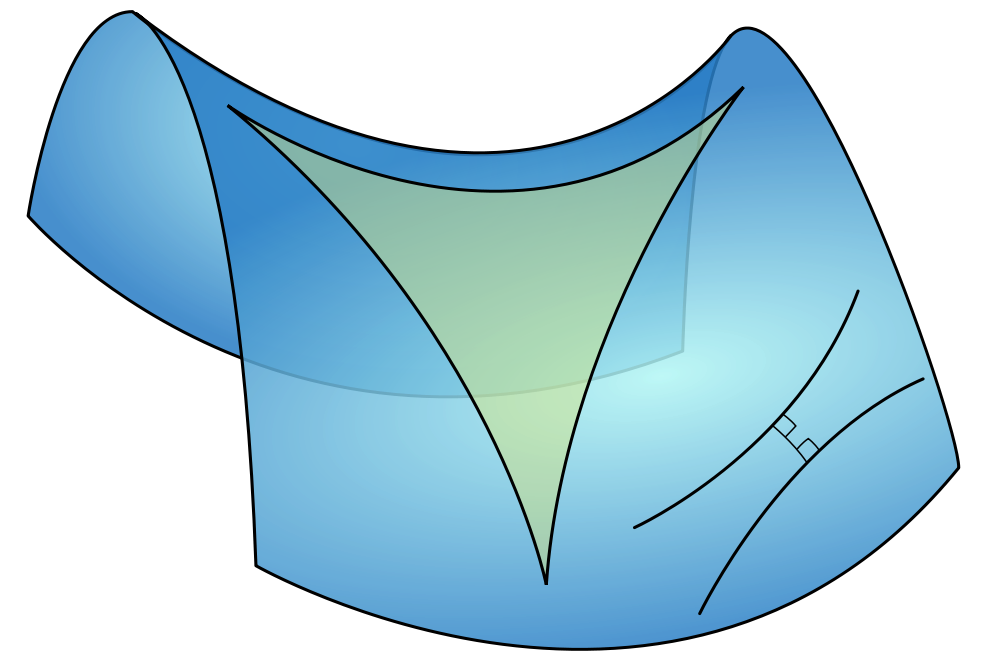}

\caption{saddle with geodesic triangle showing the negative curvature\label{fig:saddle-with-geodesic}}

\end{figure}

\emph{Step 3}: Therefore we have to understand which submanifold is
the source of the negative curvature. In the construction of the exotic
$S^{3}\times\mathbb{R}$ (see subsection \ref{sub:Constructing-the-exotic})
we used a non-trivial Casson handle. A Casson handle is an infinite
construction of failures to embed a disk best described by a tree:
the vertex is a kinky handle (a thicken disk with self-intersections)
with branches (the parts away from the self-intersection) representing
the branches. A schematic picture of the first three stages can be
found in Fig. \ref{fig:schematic-picture-CH}. 
 Alternatively, a Casson handle can be described by a tree of 1-/2-handle
pairs (see \cite{GomSti:1999}). A 1- and a 2-handle is represented
locally by the Morse functions, respectively,
\[
x_{0}^{2}+x_{1}^{2}+x_{2}^{2}-x_{3}^{2}\qquad x_{0}^{2}+x_{1}^{2}-x_{2}^{2}-x_{3}^{2}\,,
\]
i.e. by saddles having negative curvature (see Fig. \ref{fig:saddle-with-geodesic}).
 Now, the idea of the calculation is the sum over these negative curvature
using the local relation (\ref{eq:scale-quadratic-expansion}) to
arrive at a formula for $a(t)$ during the inflation process.

\emph{Step 4}: By using the tree of the Casson handle, we obtain a
countable infinite sum of contributions for (\ref{eq:scale-quadratic-expansion}).
Before we start we will clarify the geometry of the Casson handle.
The discussion of the Morse functions above uncovers the hyperbolic
geometry of the Casson handle. Therefore the tree corresponding to
the Casson handle must be interpreted as a metric tree with hyperbolic
structure in $\mathbb{H}^{2}$ and metric $ds^{2}=(dx^{2}+dy^{2})/y^{2}$.
The direction of the increasing levels $n\to n+1$ is identified with
$dy^{2}$ and $dx^{2}$ is the number of edges for a fixed level with
scaling parameter $\vartheta$. The contribution of every level in
the tree is determined by the previous level best expressed in the
scaling parameter $\vartheta$. An immersed disk at level $n$ needs
at least one disk to resolve the self-intersection point. This disk
forms the level $n+1$ but this disk is contained in the previous
disk. So we obtain for $da^{2}|_{n+1}$ at level $n+1$
\[
da^{2}|_{n+1}\sim\vartheta\cdot da^{2}|_{n}
\]
up to a constant. By using the metric $ds^{2}=(dx^{2}+dy^{2})/y^{2}$
with the interpretation ($y^{2}\to n+1$, $dx^{2}\to\vartheta$) we
obtain for the change $dx^{2}/y^{2}$ along the $x-$direction (i.e.
for a fixed $y$) $\frac{\vartheta}{n+1}$. This change determines
the scaling from the level $n$ to $n+1$, i.e. 
\[
da^{2}|_{n+1}=\frac{\vartheta}{n+1}\cdot da^{2}|_{n}=\frac{\vartheta^{n+1}}{(n+1)!}\cdot da^{2}|_{0}
\]
and after the whole summation (as substitute for an integral for the
discrete values) we obtain for the relative scaling 
\begin{equation}
a^{2}=\sum_{n=0}^{\infty}\left(da^{2}|_{n}\right)=a_{0}^{2}\cdot\sum_{n=0}^{\infty}\frac{1}{n!}\vartheta^{n}=a_{0}\cdot\exp\left(\vartheta\right)=a_{0}\cdot l_{scale}\label{eq:scaling}
\end{equation}
with $da^{2}|_{0}=a_{0}^{2}$. So, if we start with the radius $a_{0}$
of $S^{3}$ and arrive at $a$ for $\Sigma$ driven by inflation then
we will get
\[
a=a_{0}\cdot\exp\left(\frac{\vartheta}{2}\right)=a_{0}\cdot\exp\left(\frac{3\cdot vol(\Sigma)}{2\cdot CS(\Sigma)}\right)\,.
\]
The formula reflects only the changes made to obtain $\Sigma$. That
is the reason for the appearance of the Chern-Simons invariant $CS(\Sigma)$
only. If we set the starting 3-sphere of Planck size $a_{0}=L_{P}$
then one obtains
\begin{equation}
a=L_{P}\cdot\exp\left(\frac{3\cdot vol(\Sigma)}{2\cdot CS(\Sigma)}\right)\,.\label{eq:inflation-scaling}
\end{equation}

\subsection{Energy, time and expansion scale\label{sub:Energy-time-scale}}

Now we will calculate the scaling factor for our inflation model above.
Therefore we choose $\Sigma=\Sigma(8_{10})$ and using the package
SnapPea by J. Weeks%
\footnote{see the link http://www.geometrygames.org/SnapPea/index.html%
} to obtain the values for $vol(\Sigma(8_{10}))$ and $CS(\Sigma(8_{10}))$:
\begin{eqnarray*}
vol(\Sigma(8_{10})) & = & 8.65115...\\
CS(\Sigma(8_{10})) & = & 0.15616...
\end{eqnarray*}
leading to the expansion factor
\[
l_{scale}=\exp\left(\frac{3\cdot vol(\Sigma)}{2\cdot CS(\Sigma)}\right)\approx\exp(83.131...)\approx1.3\cdot10^{36}\,.
\]
This factor (more than 60 e-folds) is enough to explain the homogeneity
and the flatness of the universe. For the energy scale we need the
Morse-theoretic model from subsection \ref{sub:morse-theoretic-description}.
In our model we have a simple interpretation of the potential $\phi^{4}-\phi^{2}$
for the $SU(2)-$valued scalar field $\phi$. If the $1-/2-$handle
pair can be canceled then the potential will be reduced to $\phi^{4}$
or (by a simple transformation) to $\phi^{2}$. But then there is
no symmetry-breaking phase. So, if we cancel the handle pair then
we stop the inflation. The cancellation of the handle pair represents
the energy which is needed to implement the inflation. Therefore we
have to determine the number of levels which are needed to cancel
the handle pair. In \cite{Fre:88} Freedman showed that three levels
are needed to construct a disk which cancels the handle pair. Therefore
the first three levels determine the energy of the symmetry breaking
needed for the inflation process. This argument gives the energy scaling
factor
\[
e_{scale}=\sum_{n=0}^{3}\frac{\vartheta^{n}}{n!}=1+\frac{\vartheta}{1}+\frac{\vartheta^{2}}{4}+\frac{\vartheta^{3}}{6}
\]
reducing the energy $E_{0}$ (before the symmetry breaking) to
\[
E_{scale}=\frac{E_{0}}{e_{scale}}\,.
\]
For the scaling 
\[
\vartheta=\frac{3\cdot vol(\Sigma)}{2\cdot CS(\Sigma)}\approx83.131....
\]
we obtain the energy scaling factor

\begin{equation}
e_{scale}=1+\frac{\vartheta}{1}+\frac{\vartheta^{2}}{2}+\frac{\vartheta^{3}}{6}\approx115172.2606\label{eq:energy-scale}
\end{equation}
as a sum of the first three contributions of the Casson handle. Assuming
the Planck energy at the beginning $E_{0}=E_{Planck}$, then the energy
\begin{equation}
E_{scale}=\frac{E_{Planck}}{e_{scale}}\approx1.0591\cdot10^{14}\: GeV\label{eq:energy-fraction}
\end{equation}
is obtained. For the determination of the time scale, we have to look
for the periodic structure in the Casson handle. By using Freedmans
reembedding theorems \cite{Fre:82}, we know that a Casson handle
consists of a 5-level substructures which build a Casson handle%
\footnote{Freedman parametrizes all Casson handle using a binary tree where
each vertex represents a 5-level of the Casson handle.%
}. Therefore the periodic structure is given by these 5-levels (the
''atoms'' of the Casson handle). Then we obtain for the time scale

\[
t_{scale}=\sum_{n=0}^{5}\frac{\vartheta^{n}}{n!}\approx3.5\cdot10^{7}
\]
and if the inflation starts at Planck time
\[
T_{scale}=T_{Planck}\cdot t_{scale}\approx3\cdot10^{-36}s
\]
which also agrees with the current measurements.

\section{Reheating\label{sec:Reheating}}

As explained in the previous sections, if there is an transition from
a spherical geometry ($k=+1$) to a hyperbolic geometry ($k=-1$)
then there is a period of exponential growth. After this period there
is the usual expansion. But now we have a problem: hyperbolic 3-manifolds
do not scale! This property, also called Mostow rigidity (see subsection
\ref{sub:Mostow-rigidity}), is central for hyperbolic manifolds.
In particular, the volume of hyperbolic 3-manifolds is a topological
invariant. Above we used this property to explain why inflation ends
after a finite time. So, after the formation of the hyperbolic 3-manifold
there is no further expansion. But there is a simple way to overcome
this difficulty. Let $\Sigma$ be the hyperbolic homology 3-sphere.
The 3-manifold 
\[
\Sigma\#P=(\Sigma\setminus D^{3})\cup_{S^{2}}(P\setminus D^{3})
\]
as connected sum of $\Sigma$ and a non-hyperbolic homology sphere
$P$ is a solution to this problem. The hyperbolic 3-manifold $\Sigma$
cannot scale (by Mostow rigidity) but the other homology 3-sphere
$P$ can change its size. The usage of a 3-sphere instead of $P$
is also excluded because $\Sigma\#S^{3}=\Sigma$. If the curvature
of $\Sigma\#P$ is dominated by the negative curvature of the hyperbolic
3-manifold $\Sigma$ then we obtain also the inflation as described
above. Therefore in a realistic theory of geometric inflation we need
a more complex 3-manifold. But here we will stop making these remarks
and go over to the reheating.

\subsection{Matter coupling\label{sub:Matter-coupling}}

The inflation, as a process to form $\Sigma$, will stop after the
formation of $\Sigma$ (as transition from the 3-sphere $S^{3}$).
Therefore we have to understand the meaning of $\Sigma$. The appearance
of $\Sigma$ in $S^{3}\times_{\theta}\mathbb{R}$ is a sign for the
exotic smoothness structure. In \foreignlanguage{english}{\cite{AsselmeyerRose2012}}
we studied the effect of exotic smoothness to the Einstein-Hilbert
action. The main claim was the derivation of the Dirac and the Yang-Mills
action using the properties of the exotic smoothness structure. In
particular we identified special 3-manifolds with fermions and bosons:
\begin{enumerate}
\item Fermions: complements $S^{3}\setminus(K\times D^{2})$ of a knot $K$,
the complement is a hyperbolic 3-manifold with boundary a torus
\item Bosons: vector (gauge) bosons as torus bundles ($T^{2}\times[0,1]$
as local structure, no hyperbolic geometry).
\end{enumerate}
We will shortly introduce the model now applied to our $S^{3}\times_{\theta}\mathbb{R}$.
In subsection \ref{sub:foliation-of-exotic-S3xR} we described the
foliation of $S^{3}\times_{\theta}\mathbb{R}$. In particular we obtained
a cobordism $W(L_{\Sigma})$ between $D^{2}\times L_{\Sigma}$ and
$\bigcup_{n}D^{2}\times S^{1}$. $n-1$ components of the link $L_{\Sigma}$
are trivial and not linked with each other. Therefore the cobordism
$W(L_{\Sigma})$ reduces to a cobordism $W(K)$ between $N(K)=K\times D^{2}$
and $S^{1}\times D^{2}$ (solid torus). In section \ref{sec:FLR-cosmology},
we used the Einstein equation (\ref{eq:Einstein-equation}) to derive
some properties of the model. But this equation is obtained from the
Einstein-Hilbert action and we considered the restriction
\[
S_{EH}(W(K))=\intop_{W(K)}R\,\sqrt{g}d^{4}x
\]
to the cobordism $W(K)$. Furthermore we restrict the action to a
small neighborhood $N(K)\times[0,1]$ of the boundary $N(K)$ using
a product metric

\begin{equation}
ds^{2}=d\theta^{2}+h_{ik}dx^{i}dx^{k}\label{eq:product-metric}
\end{equation}
with coordinate $\theta$ on $[0,1]$ and metric $h_{ik}$ on $N(K)$.
By the ADM formalism with the lapse $N$ and shift function $N^{i}$
one gets a relation between the 4-dimensional $R$ and the 3-dimensional
scalar curvature $R_{(3)}$ (see \cite{MiThWh:73} (21.86) p. 520)
\begin{equation}
\sqrt{g}\, R\: d^{4}x=N\sqrt{h}\:\left(R_{(3)}+||n||^{2}((tr\mathbf{K})^{2}-tr\mathbf{K}^{2})\right)d\theta\, d^{3}x\label{eq:ADM-splitting}
\end{equation}
with the normal vector $n$ and the extrinsic curvature $\mathbf{K}$.
The embedding $N(K)\hookrightarrow N(K)\times[0,1]$ can be chosen
so that $\mathbf{K}=const.$ Then we obtain the action
\[
S_{EH}(N(K)\times[0,1])=\intop_{N(K)\times[0,1]}R\sqrt{g}d^{4}x=\intop_{N(K)}R_{(3)}\sqrt{h}N\, d^{3}x=S_{EH}(N(K))\,.
\]
The integral over $N(K)=K\times D^{2}$ is completely determined by
the boundary (the disk is flatly embedded). It can be calculated as
a term over the boundary $\partial N(K)=K\times S^{1}$, a knotted
torus, i.e. we obtain 
\begin{eqnarray*}
S_{EH}(N(K)) & = & \intop_{N(K)}R_{(3)}\sqrt{h}N\, d^{3}x=\intop_{\partial(N(K))}X\sqrt{h}d^{2}x\\
 & = & S_{EH}(\partial(N(K)))
\end{eqnarray*}
where $X$ is a 2-dimensional expression for the boundary term of
the Einstein-Hilbert action. We will use the same symbol for the 2-dimensional
metric $h$ and its restriction to the boundary submanifold. Now we
are looking for the action at the boundary. As shown by York \cite{York1972},
the fixing of the conformal class of the spatial metric in the ADM
formalism leads to a boundary term which can be also found in the
work of Hawking and Gibbons \cite{GibHaw1977}. Also Ashtekar et.al.
\cite{Ashtekar08,Ashtekar08a} discussed the boundary term in the
Palatini formalism. The main reason for the introduction of the boundary
term came from the Hamiltonian formulation of Einsteins theory. It
has been known since the 1960\textquoteright{}s (see \cite{MiThWh:73}
section 21.4-21.8) that in the Hamiltonian quantization of gravity
it is essential to include boundary terms in the action, as this allows
to define consistently the momentum conjugate to the metric. This
makes it necessary to modify the Einstein-Hilbert action by adding
to it a surface integral term so that the variation of the action
becomes well defined and yields the Einstein field equations. All
these discussions enforce us to choose the following action term at
the boundary $\partial(N(K))$ 
\begin{equation}
S_{EH}(\partial(N(K)))=\intop_{\partial(N(K))}H_{\partial}\:\sqrt{h}d^{2}x\label{eq:action2Dfermi}
\end{equation}
with $H_{\partial}$ as \emph{mean curvature} of $\partial(N(K))$,
i.e. the trace of the second fundamental form. If we consider $\partial N(K)\times[0,1]$
then $H_{\partial}$ is constant along $[0,1]$ and we obtain the
integral relation 
\begin{equation}
S_{EH}(N(K)\times[0,1])=S_{EH}(\partial N(K)\times[0,1])=\intop_{\partial(N(K))\times[0,1]}H_{\partial}\:\sqrt{h}d^{2}x\, d\theta\label{eq:action-fermi}
\end{equation}
using the product metric (\ref{eq:product-metric}) and the splitting
(\ref{eq:ADM-splitting}). The action (\ref{eq:action-fermi}) above
is completely determined by the knotted torus $\partial N(K)=K\times S^{1}$
and its mean curvature $H_{\partial N(K)}$. This knotted torus is
an immersion of a torus $S^{1}\times S^{1}$ into $\mathbb{R}^{3}$.
The well-known \emph{Weierstrass representation} can be used to describe
this immersion. As proved in \cite{SpinorRep1996,Friedrich1998} there
is an equivalent representation via spinors. This so-called \emph{Spin
representation} of a surface gives back an expression for $H_{\partial N(K)}$
and the Dirac equation as geometric condition on the immersion of
the surface. As we will show below, the term (\ref{eq:action-fermi})
can be interpreted as Dirac action of a spinor field. Here we have
an immersion of a torus $I:T^{2}=S^{1}\times S^{1}\to\mathbb{R}^{3}$
with image the knotted torus $im(I)=T(K)=K\times S^{1}$ that is the
boundary $\partial N(K)$ of $N(K)$. This immersion $I$ can be defined
by a spinor $\varphi$ on $T^{2}$ fulfilling the Dirac equation
\begin{equation}
D\varphi=H\varphi\label{eq:2D-Dirac}
\end{equation}
with $|\varphi|^{2}=1$ (or an arbitrary constant) (see Theorem 1
of \cite{Friedrich1998}). As discussed above a spinor bundle over
a surface splits into two sub-bundles $S=S^{+}\oplus S^{-}$ with
the corresponding splitting of the spinor $\varphi$ in components
\[
\varphi=\left(\begin{array}{c}
\varphi^{+}\\
\varphi^{-}
\end{array}\right)
\]
and we have the Dirac equation
\[
D\varphi=\left(\begin{array}{cc}
0 & \partial_{z}\\
\partial_{\bar{z}} & 0
\end{array}\right)\left(\begin{array}{c}
\varphi^{+}\\
\varphi^{-}
\end{array}\right)=H\left(\begin{array}{c}
\varphi^{+}\\
\varphi^{-}
\end{array}\right)
\]
with respect to the coordinates $(z,\bar{z})$ on $T^{2}$. 

In dimension 3, the spinor bundle has the same fiber dimension as
the spinor bundle $S$ (but without a splitting $S=S^{+}\oplus S^{-}$into
two sub-bundles). Now we define the extended spinor $\phi$ over $T^{2}\times[0,1]=S^{1}\times S^{1}\times[0,1]$
via the restriction $\phi|_{T^{2}}=\varphi$. The spinor $\phi$ is
constant along the normal vector $\partial_{N}\phi=0$ fulfilling
the 3-dimensional Dirac equation
\begin{equation}
D^{3D}\phi=\left(\begin{array}{cc}
\partial_{N} & \partial_{z}\\
\partial_{\bar{z}} & -\partial_{N}
\end{array}\right)\phi=H\phi\label{eq:Dirac-equation-3D}
\end{equation}
induced from the Dirac equation (\ref{eq:2D-Dirac}) via restriction
and where $|\phi|^{2}=const.$ Especially one obtains for the mean
curvature of $\partial N(K)\times[0,1]=K\times S^{1}\times[0,1]$
(up to a constant from $|\phi|^{2}$)
\begin{equation}
H=\bar{\phi}D^{3D}\phi\,.\label{eq:mean-curvature-3D}
\end{equation}
This mean curvature can be put into the action (\ref{eq:action-fermi})
to obtain the 3-dimensional Dirac action
\begin{equation}
S_{EH}(\partial N(K)\times[0,1]=\intop_{[0,1]\times\partial N(K)}\bar{\phi}D^{3D}\phi\:\sqrt{g}\, d\theta d^{2}x\,.\label{eq:3D-action-fermion}
\end{equation}
For the extension of this action to 4 dimensions, we consider a slightly
more general case. Let $\iota:\Sigma\hookrightarrow M$ be an immersion
of the 3-manifold $\Sigma$ into the 4-manifold $M$ with the normal
vector $\vec{N}$. The spin bundle $S_{M}$ of the 4-manifold splits
into two sub-bundles $S_{M}^{\pm}$ where one subbundle, say $S_{M}^{+},$
can be related to the spin bundle $S_{\Sigma}$ of the 3-manifold.
Then the spin bundles are related by $S_{\Sigma}=\iota^{*}S_{M}^{+}$
with the same relation $\phi=\iota_{*}\Phi$ for the spinors ($\phi\in\Gamma(S_{\Sigma})$
and $\Phi\in\Gamma(S_{M}^{+})$). Let $\nabla_{X}^{M},\nabla_{X}^{\Sigma}$
be the covariant derivatives in the spin bundles along a vector field
$X$ as section of the bundle $T\Sigma$. Then we have the formula
\begin{equation}
\nabla_{X}^{M}(\Phi)=\nabla_{X}^{\Sigma}\phi-\frac{1}{2}(\nabla_{X}\vec{N})\cdot\vec{N}\cdot\phi\label{eq:covariant-derivative-immersion}
\end{equation}
with the obvious embedding $\phi\mapsto\left(\begin{array}{c}
\phi\\
0
\end{array}\right)=\Phi$ of the spinor spaces. The expression $\nabla_{X}\vec{N}$ is the
second fundamental form of the immersion where the trace $tr(\nabla_{X}\vec{N})=2H$
is related to the mean curvature $H$. Then from (\ref{eq:covariant-derivative-immersion})
one obtains a similar relation between the corresponding Dirac operators
\begin{equation}
D^{4D}\Phi=D^{3D}\phi-H\phi\label{eq:relation-Dirac-3D-4D}
\end{equation}
with the Dirac operator $D^{3D}$ defined via (\ref{eq:Dirac-equation-3D}).
Together with equation (\ref{eq:Dirac-equation-3D}) we obtain
\begin{equation}
D^{4D}\Phi=0\label{eq:Dirac-equation-4D}
\end{equation}
i.e. $\Phi$ is a parallel spinor. This Dirac equation is obtained
by varying the action
\begin{equation}
\delta\intop_{M}\bar{\Phi}D^{4D}\Phi\sqrt{g}\: d^{4}x=0\label{eq:4D-variation}
\end{equation}
Importantly, this variation has a different interpretation in contrast
to varying the 3-dimensional action. Both variations look very similar.
But in (\ref{eq:4D-variation}) we vary over smooth maps $\Sigma\to M$
which are not conformal immersions (i.e. represented by spinors $\Phi$
with $D^{4D}\Phi\not=0$). Only the choice of the extremal action
selects the conformal immersion among other smooth maps. Especially
the spinor $\Phi$ (as solution of the 4-dimensional Dirac equation)
is localized at the immersed 3-manifold $\Sigma$. The 3-manifold
$\Sigma$ moves along the normal vector (see the relation (\ref{eq:covariant-derivative-immersion})
between the covariant derivatives representing a parallel transport).
Therefore the 3-dimensional action (\ref{eq:3D-action-fermion}) can
be extended to the whole 4-manifold (but for a spinor $\Phi$ of fixed
chirality). Especially we have a unique fermionic action on the manifold
$M$.

Applied to our example, we obtain for the Einstein-Hilbert action
\[
S_{EH}(N(K)\times[0,1])=\intop_{N(K)\times[0,1]}\bar{\Phi}D^{4D}\Phi\sqrt{g}\: d^{4}x
\]
the 4-dimensional Dirac action. But then we have the close relation
between the knot 
\[
\mbox{Knot }K=8_{10}\rightleftharpoons\mbox{Spinor }\Phi
\]
forming our result\\
\emph{The formation of the hyperbolic homology 3-sphere $\Sigma$
produces spinorial matter.}\\
But we may ask: Is this kind of matter the correct baryonic matter
in the cosmological sense? By using the hyperbolic structure on $\Sigma$
we can simple claim: Yes. Hyperbolic 3-manifolds do not scale by Mostow
rigidity (see subsection \ref{sub:Mostow-rigidity}). So, if we consider
the model $\Sigma\#P$ above, then the size of $P$ can be increase
whereas $\Sigma$ remains constant. Therefore an energy density associated
to $\Sigma$ scales like $a^{-3}$ if $a$ is the scale of $P$. We
obtain the correct scaling behavior of matter in cosmology. The appearance
of interactions demand a more complex knot and we refer to our paper
\cite{AsselmeyerRose2012} for the details.

\subsection{Supercooled expansion}

For a discussion of the temperature, we have to consider the degrees
of freedom. In a topological model, one usually considers the topological
degrees, i.e. the topology-determing submanifolds. In case of 3-manifolds,
the fundamental group is the main structure. For $S^{3}\to\Sigma=\Sigma(8_{10})$,
we have the transition of the trivial fundamental group $\pi_{1}(S^{3})=0$
(containing only the unit element, a contractable closed curve) to
the infinite fundamental group
\begin{eqnarray*}
\pi_{1}(\Sigma(8_{10})) & = & \langle a,b,c,d,f|\, d=adfc\,;\, b^{-1}=adb^{-1}af=df^{-1}d\,;\\
 &  & \, ac^{.1}a^{-1}c=f^{-1}\,;\, a^{2}d=ca^{-1}bafbdca^{-1}c\rangle
\end{eqnarray*}
with five generators (and four relations). The new degrees of freedom
are cause the lowering of the temperature during the inflation phase.
We obtain a supercooled expansion. The correct temperature value is
directly related to the energy scale (\ref{eq:energy-scale}) by the
Boltzmann constant $k_{B}$. Explicitly we have the energy fraction
\[
\frac{E_{0}}{E_{1}}=e_{scale}
\]
between the energy $E_{0}$ at the beginning and the energy $E_{1}$
at the end of the inflation. But this formula is identical to (\ref{eq:energy-fraction})
and we obtain the temperature fraction
\[
\frac{E_{0}}{E_{1}}=\frac{T_{0}}{T_{1}}=e_{scale}=1+\vartheta+\frac{\vartheta^{2}}{4}+\frac{\vartheta^{3}}{6}\approx115172.2606
\]
 using (\ref{eq:energy-scale}), $E=k_{B}\cdot T$ and the value for
$\vartheta\approx83.131..$. Then we obtain the decrease of the temperature
of about 115000 K (from the Planck temperature $10^{32}\, K$ to $10^{27}\, K$)
in agreement with the usual inflation models.

\subsection{Reheating}

At the end of the inflation, the hyperbolic homology 3-sphere $\Sigma$
is formed. But by the results of subsection \ref{sub:Matter-coupling},
matter is created now. The inflation phase ends and the energy is
now transferred to the matter. For an explicit construction of the
dynamics, we have to use our Morse-theoretic model of subsection \ref{sub:morse-theoretic-description}.
In the process to cancel the handle pair we have to introduce a Casson
handle. We learned above that 3-levels of the Casson handle are enough
to cancel the handle pair but at the same time new 1-/2-handle pairs
appear. For the following argumentation we consider the simplest Casson
handle (represented by the unbranched tree $T_{0}$), i.e. an unbranched
chain of 1-/2-handle pairs. As explained in subsection \ref{sub:morse-theoretic-description},
the cancellation of the handle pair is equivalent to the potential
$V(\phi)=\phi^{2}$ but the Casson handle introduces new (infinite
many) 1-/2-handle pairs. From the Morse-theoretic point of view, these
handles are attached to the critical point of $\phi^{2}$, i.e. to
$\phi=0$. The Morse function $\Psi$ in the commutative diagram (\ref{eq:commuting-diagram-1})
must be modified for the Casson handle $CH$. Now we have to consider
a Morse function $\Psi':\Sigma\cup CH\to\mathbb{R}$. The Casson handle
is an periodic structure \cite{Kato2004} (or more mathematically:
an end-periodic manifold) and we need an extra parameter for the Morse
function reflecting the period. We model the handle pairs by a periodic
function like sine or cosine. So, the Morse function for $CH$ is
$\sin(k)$ with the new parameter $k$. The attachment to the critical
value $\phi=0$ can be realized by the product $\phi^{2}\cdot\sin(k)$.
Then it modifies the whole potential to 
\[
V(\phi)=\phi^{2}+\phi^{2}\cdot\sin(k)
\]
and we obtain the new Lagrangian b modifying (\ref{eq:Lagrangian-scalar-field})
\begin{equation}
L_{\phi}=D_{\mu}\phi\cdot D^{\mu}\phi+A\cdot\phi^{2}+B\cdot\phi^{2}\cdot\sin(k)\,.\label{eq:Lagrangian-parametric-resonance}
\end{equation}
But we should keep one thing in mind: in the derivation of the expansion
rate in subsection \ref{sub:Curvature-contribution-of-CH} we choose
time as the direction of the growing tree. Therefore the parameter
$k=C\cdot t$ is proportional to time $t$. In the following we consider
the time-dependent changes of the scalar field, the fluctuations,
denoted by $\chi(t)$. For these fluctuations we obtain the Lagrangian
\[
L_{\chi}=\dot{\chi}^{2}+\chi^{2}(A+B\cdot\sin(C\cdot t))
\]
and the equation (the dot is the time-derivative)
\[
\ddot{\chi}+\chi(A+B\cdot\sin(C\cdot t))=0
\]
also called Mathieu equation (for parametric resonance). At least
for the fluctuation $\chi$, we obtain also the model of parametric
resonance as the basic process for reheating and matter creation. 

Currently, the model is not fully realistic. It shows the main features
of an inflation model but it do not contain the right particle types
(like leptons or quarks). But in a forthcoming work we will address
this question further.

\subsection{Inhomogeneities}

At the end of the previous subsection, we emphasized that our model
is only a simplification. For a more realistic model, we need a more
complex 3-manifold consisting on a connected sum
\[
\Sigma=\Sigma_{1}\#\Sigma_{2}\#\cdots
\]
of homology 3-spheres. In this paper we showed that hyperbolic 3-manifolds
induces inflation. But a critical look into the arguments of the derivation
uncovers a different possibility. A hyperbolic 3-manifold is (uniquely)
characterized by the fact that every plane in a hyperbolic 3-manifolds
has a negative curvature%
\footnote{Sometimes one claimed that every ''direction'' has a negative curvature.%
}. In subsection \ref{sub:Mostow-rigidity} we also described the other
3-geometries. In the list, one can find geometries like $\mathbb{H}^{2}\times\mathbb{R}$,
$\tilde{SL}_{2}$ and SOL with negative curvature along some plane.
These geometries are also able to generate an accelerated expansion
but in a weaker sense. Therefore a sum of (geometrically) different
homology 3-spheres induces an inhomogeneous inflation process. But
we will study it in a forthcoming paper more carefully.

\subsection{Dependence on initial conditions and Quantum fluctuations\label{sub:Dependence-on-initial-condition-quantum-fluctuation}}

One problem of the current inflation theory is the choice of the initial
conditions: only a tiny class of conditions show inflation (see Penrose
\cite{Penrose1989}). In contrast to this discussion of usual inflation,
our model is more robust. The cause of the inflation is a topological
transition explaining the robustness of the process. Indeed, the model
is independent of the concrete smoothness structure. We have to demand
only that the smoothness structure is not the standard one. But the
richness of exotic smoothness structures imply that the choice of
the standard smoothness structure is of measure zero. Secondly we
need a homology 3-sphere with negative curvature. As one learns from
Thurstons work, most of the geometric structures on 3-manifolds are
hyperbolic structures. In particular, there is only homology 3-sphere
not diffeomorphic to the 3-sphere which does carry a geometric structure
of positive curvature: the Poincare sphere. Again, only a small finite
number among infinite configurations do not lead to inflation. Therefore
our model has a converse behavior: nearly all initial conditions are
producing an inflationary scenario.

In the inflation model, quantum fluctuations are amplified to be the
cause of structure formation in the cosmos. Therefore we need a counterpart
of the quantum fluctuations in our geometric model. In \cite{AsselmeyerKrol2011c,AsselmeyerKrol2013},
we constructed a geometric quantum state by using a wild embedding.
Let $i:K\to M$ be an embedding, i.e. $i(K)$ is homeomorphic to $K$.
An embedding is called wild, if $i(K)$ is an infinite polyhedron
(a triangulation with infinite many triangles) which can be never
reduced to a finite polyhedron. Otherwise, an embedding is called
tame. Famous examples of wild embeddings are Alexanders horned sphere
and Antoines necklace. In \cite{AsselmeyerKrol2013}, we showed that
wild embeddings are a geometric expression for a quantum state (as
Hilbert space vector in an infinite-dimensional Hilbert space). The
transition of a wild embedding to a tame embedding is the transition
of a quantum state to a classical state. The wild embedding can be
also described by a fractal which is self-similar, i.e. scale-invariant.
In our exotic $S^{3}\times_{\theta}\mathbb{R}$, there is no tame
embedded $S^{3}$, i.e. every 3-sphere is wildly embedded. In particular,
the 3-sphere in the cobordism between $S^{3}$ and $\Sigma$ is also
wildly embedded. In our model, the cosmos started in a quantum state
given by a wildly embedded $S^{3}$. The quantum fluctuations are
the parts of the wildly embedded $S^{3}$. We close this paper with
these remarks which should be further explored in the forthcoming
work.

\section{Conclusion}

In this paper we developed the theory of \emph{geometric inflation}.
According to this model, an inflationary phase in the cosmic evolution
is caused by the exotic smoothness structure of our spacetime. The
exotic smoothness structure is constructed by a hyperbolic homology
3-sphere $\Sigma$. The exponential expansion has its origin in the
hyperbolic structure of the spacetime. This expansion is determined
by a single parameter $\vartheta$, the fraction of two topological
invariants for the hyperbolic homology 3-sphere: the volume and the
Chern-Simons invariant. Furthermore we obtain expressions for all
relevant quantities like energy, time and temperature depending only
on the parameter $\vartheta$. The coupling to matter can be also
expressed geometrically (using the spinor representation of embedded
surfaces). Then the reheating process has also a geometrical counterpart
and we obtain naturally the model of parametric resonance. Finally
we discuss a geometric interpretation of quantum fluctuations. One
question remains: But what is about the inflation without quantum
effects? Fortunately, there is growing evidence that the differential
structures constructed above (i.e. exotic smoothness in dimension
4) is directly related to quantum gravitational effects \cite{Ass2010,Duston2010}.
Maybe we touch only the tip of the iceberg. 
\begin{acknowledgments}
At first we want to express our gratitude to C.H. Brans, R. Gompf
and H. Rose for numerous discussions. The critical remarks of the
referees for the first version of this work is also acknowledged.
\end{acknowledgments}
\appendix

\section{Connected and boundary-connected sum of manifolds\label{sec:Connected-and-boundary-connected}}

Now we will define the connected sum $\#$ and the boundary connected
sum $\natural$ of manifolds. Let $M,N$ be two $n$-manifolds with
boundaries $\partial M,\partial N$. The \emph{connected sum} $M\#N$
is the procedure of cutting out a disk $D^{n}$ from the interior
$int(M)\setminus D^{n}$ and $int(N)\setminus D^{n}$ with the boundaries
$S^{n-1}\sqcup\partial M$ and $S^{n-1}\sqcup\partial N$, respectively,
and gluing them together along the common boundary component $S^{n-1}$.
The boundary $\partial(M\#N)=\partial M\sqcup\partial N$ is the disjoint
sum of the boundaries $\partial M,\partial N$. The \emph{boundary
connected sum} $M\natural N$ is the procedure of cutting out a disk
$D^{n-1}$ from the boundary $\partial M\setminus D^{n-1}$ and $\partial N\setminus D^{n-1}$
and gluing them together along $S^{n-2}$ of the boundary. Then the
boundary of this sum $M\natural N$ is the connected sum $\partial(M\natural N)=\partial M\#\partial N$
of the boundaries $\partial M,\partial N$.

\section{Casson handle\label{sec:Casson-handle}}

Above we have seen the determination of the smoothness structure for
the compact 4-manifold $M$ by the Akbulut cork $A$, a contractable
4-manifold with boundary a homology 3-sphere, and by the involution
$\tau:\,\partial A\to\partial(M\setminus A)$. So, why we need the
Casson handle? Lets start again with a h-cobordism $W$ between $M_{0}$
and $M$. As shown by Freedman \cite{Fre:82} $W$ is topologically
trivial, i.e. $W$ is homeomorphic to $M_{0}\times[0,1]$. Thus, as
explained above, one can cancel the $2-/3-$handle pairs by using
the Whitney trick, i.e. one is able to embed a special 2-disk $D^{2}$
or equivalently a 2-handle $D^{2}\times D^{2}$. Usually one fails
in dimension four to do it but Casson \cite{Cas:73} found an infinite
construction, the Casson handle, homotopic to $D^{2}\times\mathbb{R}^{2}$.
Via a ``tour de force'' Freedman \cite{Fre:82} proved that every
Casson handle is homeomorphic to $D^{2}\times\mathbb{R}^{2}$, the
open 2-handle, leading to the topological triviality of the h-cobordism
$W$. All these constructions are relative to the boundary $\partial D^{2}\times\mathbb{R}^{2}$,
i.e. the attachment of the Casson handle is extremely important. But
by \cite{Don:87,Don:90} $W$ is in general not diffeomorphic to $M_{0}\times[0,1]$.
The reason is very simple: the Casson handle is not diffeomorphic
to $D^{2}\times\mathbb{R}^{2}$ (relative to the boundary $\partial D^{2}\times\mathbb{R}^{2}$)
\cite{Gom:84,Gom:89}. As explained above, this Casson handle determines
the involution $\tau$ of the boundary $\partial A$ and therefore
the embedding of the Akbulut cork $A$. 

Lets consider now the basic construction of the Casson handle $CH$.
Let $M$ be a smooth, compact, simple-connected 4-manifold and $f:D^{2}\to M$
a (codimension-2) mapping. By using diffeomorphisms of $D^{2}$ and
$M$, one can deform the mapping $f$ to get an immersion (i.e. injective
differential) generically with only double points (i.e. $\#|f^{-1}(f(x))|=2$)
as singularities \cite{GolGui:73}. But to incorporate the generic
location of the disk, one is rather interesting in the mapping of
a 2-handle $D^{2}\times D^{2}$ induced by $f\times id:D^{2}\times D^{2}\to M$
from $f$. Then every double point (or self-intersection) of $f(D^{2})$
leads to self-plumbings of the 2-handle $D^{2}\times D^{2}$. A self-plumbing
is an identification of $D_{0}^{2}\times D^{2}$ with $D_{1}^{2}\times D^{2}$
where $D_{0}^{2},D_{1}^{2}\subset D^{2}$ are disjoint sub-disks of
the first factor disk%
\footnote{In complex coordinates the plumbing may be written as $(z,w)\mapsto(w,z)$
or $(z,w)\mapsto(\bar{w},\bar{z})$ creating either a positive or
negative (respectively) double point on the disk $D^{2}\times0$ (the
core).%
}. Consider the pair $(D^{2}\times D^{2},\partial D^{2}\times D^{2})$
and produce finitely many self-plumbings away from the attaching region
$\partial D^{2}\times D^{2}$ to get a kinky handle $(k,\partial^{-}k)$
where $\partial^{-}k$ denotes the attaching region of the kinky handle.
A kinky handle $(k,\partial^{-}k)$ is a one-stage tower $(T_{1},\partial^{-}T_{1})$
and an $(n+1)$-stage tower $(T_{n+1},\partial^{-}T_{n+1})$ is an
$n$-stage tower union kinky handles $\bigcup_{\ell=1}^{n}(T_{\ell},\partial^{-}T_{\ell})$
where two towers are attached along $\partial^{-}T_{\ell}$. Let $T_{n}^{-}$
be $(\mbox{interior}T_{n})\cup\partial^{-}T_{n}$ and the Casson handle
\[
CH=\bigcup_{\ell=0}T_{\ell}^{-}
\]
is the union of towers (with direct limit topology induced from the
inclusions $T_{n}\hookrightarrow T_{n+1}$). 

The main idea of the construction above is very simple: an immersed
disk (disk with self-intersections) can be deformed into an embedded
disk (disk without self-intersections) by sliding one part of the
disk along another (embedded) disk to kill the self-intersections.
Unfortunately the other disk can be immersed only. But the immersion
can be deformed to an embedding by a disk again etc. In the limit
of this process one ``shifts the self-intersections into infinity''
and obtains%
\footnote{In the proof of Freedman \cite{Fre:82}, the main complications come
from the lack of control about this process. %
} the standard open 2-handle $(D^{2}\times\mathbb{R}^{2},\partial D^{2}\times\mathbb{R}^{2})$. 

A Casson handle is specified up to (orientation preserving) diffeomorphism
(of pairs) by a labeled finitely-branching tree with base-point {*},
having all edge paths infinitely extendable away from {*}. Each edge
should be given a label $+$ or $-$. Here is the construction: tree
$\to CH$. Each vertex corresponds to a kinky handle; the self-plumbing
number of that kinky handle equals the number of branches leaving
the vertex. The sign on each branch corresponds to the sign of the
associated self plumbing. The whole process generates a tree with
infinite many levels. In principle, every tree with a finite number
of branches per level realizes a corresponding Casson handle. For
every labeled based tree $Q$, let us describe a subset $U_{Q}$ of
$D^{2}\times D^{2}$. Now we will construct a $(U_{Q},\partial D^{2}\times D^{2})$
which is diffeomorphic to the Casson handle associated to $Q$. In
$D^{2}\times D^{2}$ embed a ramified Whitehead link with one Whitehead
link component for every edge labeled by $+$ (plus) leaving {*} and
one mirror image Whitehead link component for every edge labeled by
$-$(minus) leaving {*}. Corresponding to each first level node of
$Q$ we have already found a (normally framed) solid torus embedded
in $D^{2}\times\partial D^{2}$. In each of these solid tori embed
a ramified Whitehead link, ramified according to the number of $+$
and $-$ labeled branches leaving that node. We can do that process
for every level of $Q$. Let the disjoint union of the (closed) solid
tori in the $n$th family (one solid torus for each branch at level
$n$ in $Q$) be denoted by $X_{n}$. $Q$ tells us how to construct
an infinite chain of inclusions:
\[
\ldots\subset X_{n+1}\subset X_{n}\subset X_{n-1}\subset\ldots\subset X_{1}\subset D^{2}\times\partial D^{2}
\]
and we define the Whitehead decomposition $Wh_{Q}=\bigcap_{n=1}^{\infty}X_{n}$
of $Q$. $Wh_{Q}$ is the Whitehead continuum \cite{Whitehead35}
for the simplest unbranched tree. We define $U_{Q}$ to be
\[
U_{Q}=D^{2}\times D^{2}\setminus(D^{2}\times\partial D^{2}\cup\mbox{closure}(Wh_{Q}))
\]
alternatively one can also write
\begin{equation}
U_{Q}=D^{2}\times D^{2}\setminus\mbox{cone}(Wh_{Q})\label{eq:UQ-diffeo-CH}
\end{equation}
where $\mbox{cone}()$ is the cone of a space. As Freedman (see \cite{Fre:82}
Theorem 2.2) showed $U_{Q}$ is diffeomorphic to the Casson handle
$CH_{Q}$ given by the tree $Q$. We will later use this construction
in the determination of the boundary $\partial N(A)$.

\section{Foliation, foliated cobordism and foliations of $S^{3}$ \label{sec:Foliation-foliated-cobordism}}

A codimension $k$ foliation%
\footnote{In general, the differentiability of a foliation is very important.
Here we consider the smooth case only. %
} of an $n$-manifold $M^{n}$ (see the nice overview article \cite{Law:74})
is a geometric structure which is formally defined by an atlas $\left\{ \phi_{i}:U_{i}\to M^{n}\right\} $,
with $U_{i}\subset\mathbb{R}^{n-k}\times\mathbb{R}^{k}$, such that
the transition functions have the form 
\[
\phi_{ij}(x,y)=(f(x,y),g(y)),\,\left[x\in\mathbb{R}^{n-k},y\in\mathbb{R}^{k}\right]\quad.
\]
To be more precise one has \begin{definition}Let $(M,\mathcal{A})$
be an $n-$dimensional smooth manifold (of differential structure
$\mathcal{A}$) with boundary: we take $\phi_{\lambda}:U_{\lambda}\to V_{\lambda}\subset\mathbb{R}_{+}^{n}$,
$\mathbb{R}_{+}^{n}=\left\{ (x_{1},\ldots,x_{n})\in\mathbb{R}^{n}|\: x_{n}\geq0\right\} $
for a chart $(U_{\lambda},\phi_{\lambda})\in\mathcal{A}$. Let $k$
with $0\leq k\leq n$ be an integer. Let $\mathcal{F}=\left\{ L_{\alpha}|\,\alpha\in A\right\} $
be a family of arcwise connected subsets $L_{\alpha}$ of the manifold
$M$. We say that $\mathcal{F}$ is a $k-$dimensional smooth foliation
of $M$ if it satisfies the following rules. 
\begin{enumerate}
\item $L_{\alpha}\cap L_{\beta}=\emptyset$ for $\alpha,\beta\in A$ with
$\alpha\not=\beta$. 
\item $\bigcup_{\alpha\in A}(L_{\alpha})=M$
\item Given a point $p$ in $M$, there exists a chart $(U_{\lambda},\phi_{\lambda})\in\mathcal{A}$
about $p$ such that for $L_{\alpha}$ with $U_{\lambda}\cap L_{\alpha}\not=\emptyset$,
$\alpha\in A$, each (arcwise) connected component of $\phi_{\lambda}(U_{\lambda}\cap L_{\alpha})$
is of the form 
\[
\left\{ (x_{1},\ldots,x_{n})\in\phi_{\lambda}(U_{\lambda})\,|\: x_{k+1}=c_{k+1},x_{k+2}=c_{k+2},\ldots,x_{n}=c_{n}\right\} 
\]
 where $c_{k+1},c_{k+2},\ldots,c_{n}$ are constants determined by
the (arcwise) connected component. 
\item For $p\in\partial M\cap L_{\alpha}$ we denote the boundary by $\partial\mathcal{F}=\left\{ L_{\alpha}\:|\,\alpha\in A,\: L_{\alpha}\subset\partial M\right\} $
\end{enumerate}
We call $L_{\alpha}$ a leaf of the foliation $\mathcal{F}$. The
foliation $\mathcal{F}$ is also referred to as a smooth codimension
$n-k$ foliation or a smooth foliation of codimension $n-k$. \end{definition}
Intuitively, a foliation is a pattern of $(n-k)$-dimensional stripes
- i.e., submanifolds - on $M^{n}$, called the leaves of the foliation,
which are locally well-behaved. The tangent space to the leaves of
a foliation $\mathcal{F}$ forms a vector bundle over $M^{n}$, denoted
$T\mathcal{F}$.

Now we will discuss an important equivalence relation between foliations,
cobordant foliations. \begin{definition} Let $M_{0}$ and $M_{1}$
be two closed, oriented $m$-manifolds with codimension-$q$ foliations.
Then these foliated manifolds are said to be \emph{foliated cobordant}
if there is a compact, oriented $(m+1)$-manifold with boundary $\partial W=M_{0}\sqcup\overline{M}_{1}$
and with a codimension-$q$ foliation $\mathcal{F}$ transverse to
the boundary. Every leaf $L_{\alpha}$ of the foliation $\mathcal{F}$
induces leafs $L_{\alpha}\cap\partial W$ of foliations $\mathcal{F}_{M_{0}},\mathcal{F}_{M_{1}}$on
the two components of the boundary $\partial W$.\end{definition}
The resulted foliated cobordism classes $[\mathcal{F}_{M}]$ of the
manifold $M$ form an abelian group $\mathcal{CF}_{m,q}(M)$ under
disjoint union $\sqcup$ (inverse $\overline{M}$, unit $S^{q}\times S^{m-q}$,
see \cite{Tamura1992} \S29).

\begin{figure}
\includegraphics[scale=0.15]{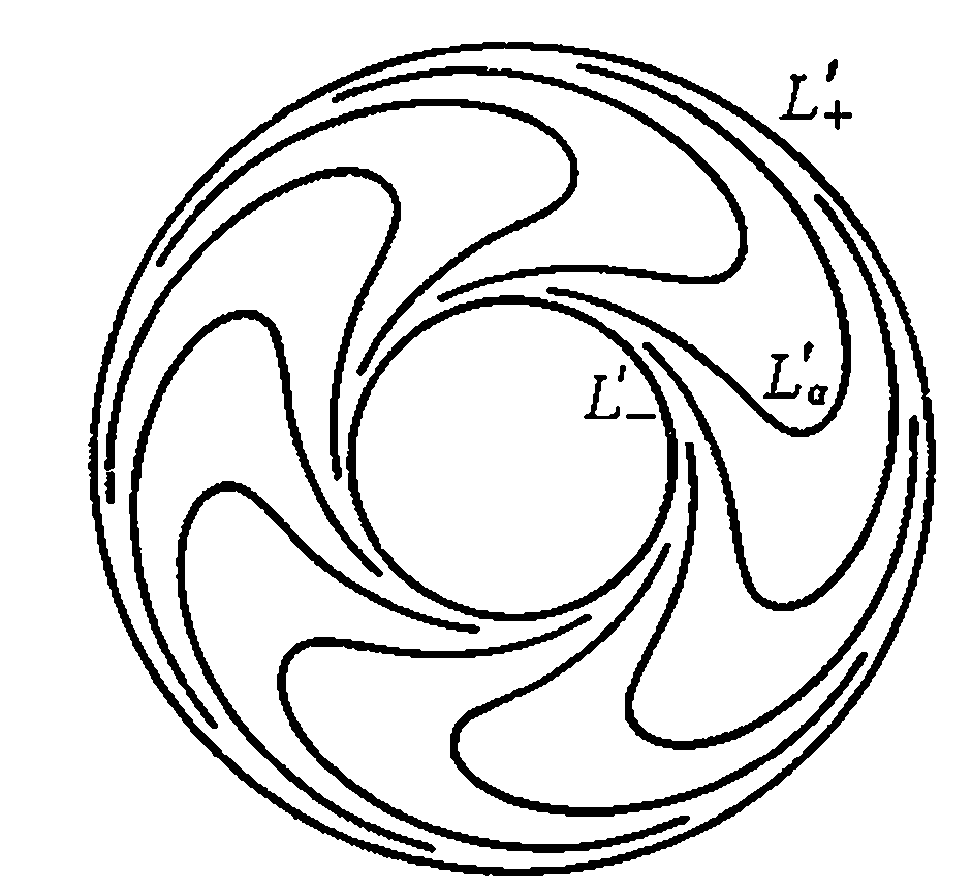}\includegraphics{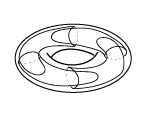}

\caption{foliation of $D^{1}\times S^{1}$(left), Reeb foliation solid torus
(right)\label{fig:foliation-torus}}
\end{figure}

One of the first examples of a nontrivial foliation is known as Reeb
foliation. Let $f:(-1,1)\to\mathbb{R}$ be a smooth function $f(t)=\exp(t^{2}/(1-t^{2}))-1$.
But every function $f$ with
\[
\begin{array}{ccc}
f(0)=0 & f(t)\geq0\qquad f(t)=f(-t)\quad & -1<t<1\\
\lim_{t\to\pm1}\frac{d^{k}}{dt^{k}}f(t)=\infty\quad & \lim_{t\to\pm1}\frac{d^{k}}{dt^{k}}\left(\frac{1}{\frac{df}{dt}(t)}\right)=0\quad & k=0,1,2,\ldots
\end{array}
\]
will also work. Define subsets $L_{\alpha}^{'},\:0\leq\alpha<1$,
and $L_{\pm}^{'}$ of $D^{1}\times S^{1}$ by
\[
L{}_{\alpha}^{'}=\left\{ t,\exp\left(2\pi(\alpha+f(t))i\right)|\:-1<t<1\right\} ,\quad L_{\pm}^{'}=\left\{ (\pm1,e^{2\pi\theta i})|\:0\leq\theta<1\right\} 
\]
defining a smooth foliation of $D^{1}\times S^{1}$(see the left Fig.
\ref{fig:foliation-torus}). 
 The join of two copies results in a smooth foliation of the torus
$T^{2}=(D^{1}\times S^{1})\cup(D^{1}\times S^{1})$. This example
can be generalized to the solid torus $D^{2}\times S^{1}$ by defining
the subsets $L_{\alpha}$ by
\[
L_{\alpha}=\left\{ x,\exp\left(2\pi(\alpha+f(|x|))i\right)|\: x\in int(D^{2})\right\} ,\:0\leq\alpha<1
\]
where $|x|$ is the distance between the origin of the disc and the
point $x$ in the interior $in(D^{2})$. The family of sets $L_{\alpha}$
and $L_{\alpha}^{'},L_{\pm}^{'}$ for the boundary $\partial(D^{2}\times S^{1})=T^{2}$
forms a smooth foliation of $D^{2}\times S^{1}$ which is known as
Reeb foliation (see the right Fig. \ref{fig:foliation-torus}).

In \cite{Thu:72} Thurston constructed a foliation of the 3-sphere
$S^{3}$ which depends on a polygon $P$ in the hyperbolic plane $\mathbb{H}^{2}$
so that two foliations are non-cobordant if the corresponding polygons
have different areas. We will present this construction now (see also
the book \cite{Tamura1992} chapter VIII for the details).

Consider the hyperbolic plane $\mathbb{H}^{2}$ and its unit tangent
bundle $T_{1}\mathbb{H}^{2}$ , i.e the tangent bundle $T\mathbb{H}^{2}$
where every vector in the fiber has norm $1$. Thus the bundle $T_{1}\mathbb{H}^{2}$
is a $S^{1}$-bundle over $\mathbb{H}^{2}$. There is a foliation
$\mathcal{F}$ of $T_{1}\mathbb{H}^{2}$ invariant under the isometries
of $\mathbb{H}^{2}$ which is induced by bundle structure and by a
family of parallel geodesics on $\mathbb{H}^{2}$. The foliation $\mathcal{F}$
is transverse to the fibers of $T_{1}\mathbb{H}^{2}$. Let $P$ be
any convex polygon in $\mathbb{H}^{2}$. We will construct a foliation
$\mathcal{F}_{P}$ of the three-sphere $S^{3}$ depending on $P$.
Let the sides of $P$ be labeled $s_{1},\ldots,s_{k}$ and let the
angles have magnitudes $\alpha_{1},\ldots,\alpha_{k}$. Let $Q$ be
the closed region bounded by $P\cup P'$, where $P'$ is the reflection
of $P$ through $s_{1}$. Let $Q_{\epsilon}$, be $Q$ minus an open
$\epsilon$-disk about each vertex. If $\pi:T_{1}\mathbb{H}^{2}\to\mathbb{H}^{2}$
is the projection of the bundle $T_{1}\mathbb{H}^{2}$, then $\pi^{-1}(Q)$
is a solid torus $Q\times S^{1}$(with edges) with foliation $\mathcal{F}_{1}$
induced from $\mathcal{F}$. For each $i$, there is an unique orientation-preserving
isometry of $\mathbb{H}^{2}$, denoted $I_{i}$, which matches $s_{i}$
point-for-point with its reflected image $s'_{i}$. We glue the cylinder
$\pi^{-1}(s_{i}\cap Q_{\epsilon})$ to the cylinder $\pi^{-1}(s'_{i}\cap Q_{\epsilon})$
by the differential $dI_{i}$ for each $i>1$, to obtain a manifold
$M=(S^{2}\setminus\left\{ \mbox{\mbox{k} punctures}\right\} )\times S^{1}$,
and a (glued) foliation $\mathcal{F}_{2}$, induced from $\mathcal{F}_{1}$.
To get a complete $S^{3}$, we have to glue-in $k$ solid tori for
the $k$ $S^{1}\times\mbox{punctures}.$ Now we choose a linear foliation
of the solid torus with slope $\alpha_{k}/\pi$ (Reeb foliation).
Finally we obtain a smooth codimension-1 foliation $\mathcal{F}_{P}$
of the 3-sphere $S^{3}$ depending on the polygon $P$.

\section{Chern-Simons invariant\label{sec:Chern-Simons-invariant}}

Let $\mathcal{P}$ be a principal $G$ bundle over the 4-manifold
$M$ with $\partial M\not=0$. Furthermore let $A$ be a connection
in $\mathcal{P}$ with the curvature 
\[
F_{A}=dA+A\wedge A
\]
and Chern class
\[
C_{2}=\frac{1}{8\pi^{2}}\int\limits _{M}tr(F_{A}\wedge F_{A})
\]
for the classification of the bundle $P$. By using the Stokes theorem
we obtain 
\begin{equation}
\int\limits _{M}tr(F_{A}\wedge F_{A})=\int\limits _{\partial M}tr(A\wedge dA+\frac{2}{3}A\wedge A\wedge A)\label{eq:stokes-CS}
\end{equation}
with the Chern-Simons invariant 
\begin{equation}
CS(\partial M,A)=\frac{1}{8\pi^{2}}\int\limits _{\partial M}tr(A\wedge dA+\frac{2}{3}A\wedge A\wedge A)\:.\label{CS-invariante}
\end{equation}
Now we consider the gauge transformation $A\rightarrow g^{-1}Ag+g^{-1}dg$
and obtain
\[
CS(\partial M,g^{-1}Ag+g^{-1}dg)=CS(\partial M,A)+k
\]
with the winding number 
\[
k=\frac{1}{24\pi^{2}}\int\limits _{\partial M}(g^{-1}dg)^{3}\in\mathbb{Z}
\]
of the map $g:M\rightarrow G$. Thus the expression 
\[
CS(\partial M,A)\bmod1
\]
is an invariant, the Chern-Simons invariant. Now we will calculate
this invariant. For that purpose we consider the functional (\ref{CS-invariante})
and its first variation vanishes 
\[
\delta CS(\partial M,A)=0
\]
because of the topological invariance. Then one obtains the equation
\[
dA+A\wedge A=0\:,
\]
i.e. the extrema of the functional are the connections of vanishing
curvature. The set of these connections up to gauge transformations
is equal to the set of homomorphisms $\pi_{1}(\partial M)\rightarrow SU(2)$
up to conjugation. Thus the calculation of the Chern-Simons invariant
reduces to the representation theory of the fundamental group into
$SU(2)$. In \cite{FinSte:90} the authors define a further invariant
\[
\tau(\Sigma)=\min\left\{ CS(\alpha)|\:\alpha:\pi_{1}(\Sigma)\rightarrow SU(2)\right\} 
\]
for the 3-manifold $\Sigma$. This invariants fulfills the relation
\[
\tau(\Sigma)=\frac{1}{8\pi^{2}}\int\limits _{\Sigma\times\mathbb{R}}tr(F_{A}\wedge F_{A})
\]
which is the minimum of the Yang-Mills action 
\[
\left|\frac{1}{8\pi^{2}}\int\limits _{\Sigma\times\mathbb{R}}tr(F_{A}\wedge F_{A})\right|\leq\frac{1}{8\pi^{2}}\int\limits _{\Sigma\times\mathbb{R}}tr(F_{A}\wedge*F_{A})
\]
i.e. the solutions of the equation $F_{A}=\pm*F_{A}$. Thus the invariant
$\tau(\Sigma)$ of $\Sigma$ corresponds to the self-dual and anti-self-dual
solutions on $\Sigma\times\mathbb{R}$, respectively. Or the invariant
$\tau(\Sigma)$ is the Chern-Simons invariant for the Levi-Civita
connection.


\begin{thebibliography}{10}
\bibitem{Anderson} M.T. Anderson. \newblock On long-time evolution in general relativity and geometrization of   3-manifolds. \newblock {\em Comm. Math. Phys.}, 222:533 -- 567, 2001.
\bibitem{Andersson} L.~Andersson. \newblock The global existence problem in general relativity. \newblock In Chrusciel and Friedrich, editors, {\em 50 Years of the {C}auchy   Problem in General Relativity}, pages 71--120, Basel, 2004. Birkhauser. \newblock arXiv:gr-qc/9911032.
\bibitem{Ashtekar08} A.~Ashtekar, J.~Engle, and D.~Sloan. \newblock Asymptotics and {H}amiltonians in a first order formalism. \newblock {\em Class. Quant. Grav.}, 25:095020, 2008. \newblock arXiv:0802.2527.
\bibitem{Ashtekar08a} A.~Ashtekar and D.~Sloan. \newblock Action and {H}amiltonians in higher dimensional general relativity:   First order framework. \newblock {\em Class.Quant.Grav.}, 25:225025, 2008. \newblock arXiv:0808.2069.
\bibitem{Ass2010} T.~Asselmeyer-Maluga. \newblock Exotic smoothness and quantum gravity. \newblock {\em Class. Q. Grav.}, 27:165002, 2010. \newblock arXiv:1003.5506.
\bibitem{Asselmeyer2007} T.~Asselmeyer-Maluga and C.H. Brans. \newblock {\em Exotic {S}moothness and {P}hysics}. \newblock WorldScientific Publ., Singapore, 2007.
\bibitem{AsselmeyerKrol2011c} T.~Asselmeyer-Maluga and J.~Kr{\'o}l. \newblock Topological quantum d-branes and wild embeddings from exotic smooth   {$R^4$}. \newblock {\em Int. J. Mod. Phys.}, {\bf A26}:3421 -- 3437, 2011. \newblock arXiv:1105.1557.
\bibitem{AsselmeyerKrol2013} T.~Asselmeyer-Maluga and J.~Kr{\'o}l. \newblock Quantum geometry and wild embeddings as quantum states. \newblock {\em Int. J. of Geometric Methods in Modern Physics}, {\bf 10}(10),   2013. \newblock will be published in Nov. 2013, arXiv:1211.3012.
\bibitem{AsselmeyerRose2012} T.~Asselmeyer-Maluga and H.~Ros{\'e}. \newblock On the geometrization of matter by exotic smoothness. \newblock {\em Gen. Rel. Grav.}, {\bf 44}:2825 -- 2856, 2012. \newblock DOI: 10.1007/s10714-012-1419-3, arXiv:1006.2230.
\bibitem{BernalSanchez2005} A.N. Bernal and M.~Sa{\'a}nchez. \newblock Smoothness of time functions and the metric splitting of globally   hyperbolic space times. \newblock {\em Commun. Math. Phys.}, 257:43--50, 2005. \newblock arXiv:gr-qc/0401112.
\bibitem{BernalSanchez2007} A.N. Bernal and M.~Sa{\'a}nchez. \newblock Globally hyperbolic spacetimes can be defined as ''causal'' instead   of ''strongly causal''. \newblock {\em Class. Quant. Grav.}, 24:745--750, 2007. \newblock arXiv:gr-qc/0611138.
\bibitem{BernalSanchez2003} A.N. Bernal and M.~S{\'a}nchez. \newblock On smooth {C}auchy hypersurfaces and {G}eroch's splitting theorem. \newblock {\em Comm. Math. Phys.}, {\bf 243}:461--470, 2003. \newblock arXiv:gr-qc/0306108.
\bibitem{BernalSanchez2006} A.N. Bernal and M.~S{\'a}nchez. \newblock Further results on the smoothability of {C}auchy hypersurfaces and   {C}auchy time functions. \newblock {\em Lett. Math. Phys.}, 183-197:{\bf 77}, 2006. \newblock gr-qc/0512095.
\bibitem{Bre:93} G.~E. Bredon. \newblock {\em Topology and Geometry}. \newblock Springer-Verlag, New York, 1993.
\bibitem{Cas:73} A.~Casson. \newblock {\em Three lectures on new infinite constructions in 4-dimensional   manifolds}, volume~62. \newblock Birkh{\"a}user, progress in mathematics edition, 1986. \newblock Notes by Lucian Guillou, first published 1973.
\bibitem{Don:83} S.~Donaldson. \newblock An application of gauge theory to the topology of 4-manifolds. \newblock {\em J. Diff. Geom.}, {\bf 18}:269--316, 1983.
\bibitem{Don:87} S.~Donaldson. \newblock Irrationality and the h-cobordism conjecture. \newblock {\em J. Diff. Geom.}, {\bf 26}:141--168, 1987.
\bibitem{Don:90} S.~Donaldson. \newblock Polynomial invariants for smooth four manifolds. \newblock {\em Topology}, {\bf 29}:257--315, 1990.
\bibitem{Dowker1997} F.~Dowker and S.~Surya. \newblock Topology change and causal continuity. \newblock {\em Phys.Rev.}, D58:124019, 1998. \newblock arXiv:gr-qc/9711070.
\bibitem{Duston2010} C.~Duston. \newblock Exotic smoothness in 4 dimensions and semiclassical {E}uclidean   quantum gravity. \newblock {\em Int.J.Geom.Meth.Mod.Phys.}, {\bf 8}:459--484, 2010. \newblock arXiv: 0911.4068.
\bibitem{Ellis1977} G.F.R. Ellis and B.G. Schmidt. \newblock Singular space-times. \newblock {\em Gen. Rel. Grav.}, 8:915--953, 1977. \newblock Review Article.
\bibitem{FinSte:90} R.~Fintushel and R.J. Stern. \newblock Instanton homology of {S}eifert fibred homology three spheres. \newblock {\em Proc. London Math. Soc.}, {\bf 61}:109--137, 1990.
\bibitem{Fre:79} M.H. Freedman. \newblock A fake {$S^3\times R$}. \newblock {\em Ann. of Math.}, {\bf 110}:177--201, 1979.
\bibitem{Fre:82} M.H. Freedman. \newblock The topology of four-dimensional manifolds. \newblock {\em J. Diff. Geom.}, {\bf 17}:357 -- 454, 1982.
\bibitem{Fre:88} M.H. Freedman. \newblock Whitehead$_3$ is a ''slice'' link. \newblock {\em Inv. Math.}, {\bf 94}:175 -- 182, 1988.
\bibitem{Friedrich1998} T.~Friedrich. \newblock On the spinor representation of surfaces in euclidean 3-space. \newblock {\em J. Geom. and Phys.}, 28:143--157, 1998. \newblock arXiv:dg-ga/9712021v1.
\bibitem{GibHaw1977} G.W. Gibbons and S.W. Hawking. \newblock Action integrals and partition functions in quantum gravity. \newblock {\em Phys. Rev. D}, 15:2752--2756, 1977.
\bibitem{GolGui:73} M.~Golubitsky and V.~Guillemin. \newblock {\em Stable {M}appings and their {S}ingularities}. \newblock Graduate Texts in Mathematics 14. Springer Verlag, New   York-Heidelberg-Berlin, 1973.
\bibitem{Gom:84} R.~Gompf. \newblock Infinite families of casson handles and topological disks. \newblock {\em Topology}, {\bf 23}:395--400, 1984.
\bibitem{Gom:89} R.~Gompf. \newblock Periodic ends and knot concordance. \newblock {\em Top. Appl.}, {\bf 32}:141--148, 1989.
\bibitem{GomSti:1999} R.E. Gompf and A.I. Stipsicz. \newblock {\em 4-manifolds and {K}irby {C}alculus}. \newblock American Mathematical Society, 1999.
\bibitem{Gordon1975} M.A. Gordan. \newblock Knots, homology spheres, and contractible 4-manifolds. \newblock {\em Topology}, {\bf 14}:151--172, 1975.
\bibitem{Guth1981} Alan~H. Guth. \newblock Inflationary universe: A possible solution to the horizon and   flatness problems. \newblock {\em Phys. Rev. D}, 23:347--356, 1981.
\bibitem{HawEll:94} S.W Hawking and G.F.R. Ellis. \newblock {\em The Large Scale Structure of Space-Time}. \newblock Cambridge University Press, 1994.
\bibitem{JacSha:79} W.~Jaco and P.~Shalen. \newblock {\em Seifert fibered spaces in 3-manifolds}, volume~{\bf 21} of {\em   Mem. Amer. Math. Soc.} \newblock AMS, 1979.
\bibitem{Kato2004} T.~Kato. \newblock {ASD} moduli space over four-manifolds with tree-like ends. \newblock {\em Geom. Top.}, 8:779 -- 830, 2004. \newblock arXiv:math.GT/0405443.
\bibitem{KirSie:77} R.~Kirby and L.C. Siebenmann. \newblock {\em Foundational essays on topological manifolds, smoothings, and   triangulations}. \newblock Ann. Math. Studies. Princeton University Press, Princeton, 1977.
\bibitem{Kir:89} R.C. Kirby. \newblock {\em The {T}opology of 4-{M}anifolds}. \newblock Lecture Notes in Mathematics. Springer Verlag, Berlin-New York, 1989.
\bibitem{WMAP-7-years} E.~Komatsu, K.~M. Smith, J.~Dunkley, C.~L. Bennett, B.~Gold, G.~Hinshaw,   N.~Jarosik, D.~Larson, M.~R. Nolta, L.~Page, D.~N. Spergel, M.~Halpern, R.~S.   Hill, A.~Kogut, M.~Limon, S.~S. Meyer, N.~Odegard, G.~S. Tucker, J.~L.   Weiland, E.~Wollack, and E.~L. Wright. \newblock Seven-year {Wilkinson Microwave Anisotropy Probe (WMAP)}   observations: Cosmological interpretation. \newblock {\em Astrophys.J.Suppl.}, 192:18, 2011.
\bibitem{SpinorRep1996} R.~Kusner and N.~Schmitt. \newblock {The} {Spinor} {Rrepresentation} {of} {Surfaces} {in} {Space}. \newblock arXiv:dg-ga/9610005v1, 1996.
\bibitem{Law:74} H.B. Lawson. \newblock Foliations. \newblock {\em BAMS}, 80:369 -- 418, 1974.
\bibitem{InflationBook} A.R. Liddle and D.H. Lyth. \newblock {\em Cosmological Inflation and Large-Scale Structure}. \newblock Cambridge University Press, Cambridge, 2000.
\bibitem{Linde1982} Andre~D. Linde. \newblock A new inflationary universe scenario: A possible solution of the   horizon, flatness, homogeneity, isotropy and primordial monopole problems. \newblock {\em Physics Letters B}, 108:389 -- 393, 1982.
\bibitem{Livingston1981} Ch. Livingston. \newblock Homology cobordisms of 3-manifolds, knot concordances, and prime   knots. \newblock {\em Pacific J. Math.}, {\bf 94}:193--206, 1981.
\bibitem{Livingston2004} Ch. Livingston. \newblock The concordance genus of knots. \newblock {\em Alg. Geom. Top.}, {\bf 4}:1--22, 2004.
\bibitem{Mil:61} J.~Milnor. \newblock A procedure for killing the homotopy groups of differentiable   manifolds. \newblock In {\em Proc. Symp. in Pure Math. 3 (Differential Geometry)}, pages   39--55. Amer. Math. Soc., 1961.
\bibitem{Mil:62} J.~Milnor. \newblock A unique decomposition theorem for 3-manifolds. \newblock {\em Amer. J. Math.}, {\bf 84}:1--7, 1962.
\bibitem{Mil:65} J.~Milnor. \newblock {\em Lectures on the h-cobordism theorem}. \newblock Princeton Univ. Press, Princeton, 1965.
\bibitem{MiThWh:73} C.~Misner, K.~Thorne, and J.~Wheeler. \newblock {\em Gravitation}. \newblock Freeman, San Francisco, 1973.
\bibitem{Mos:68} G.D. Mostow. \newblock Quasi-conformal mappings in $n$-space and the rigidity of hyperbolic   space forms. \newblock {\em Publ. Math. IH�S}, {\bf 34}:53--104, 1968.
\bibitem{Mun:60} J.~Munkres. \newblock Obstructions to the smoothing of pieceswise-differential   homeomeomorphisms. \newblock {\em Ann. Math}, 72:621--554, 1960.
\bibitem{Penrose1989} R.~Penrose. \newblock Difficulties with inflationary cosmology. \newblock {\em Annals of the New York Academy of Sciences}, {\bf 271}:249--264,   1989. \newblock doi:10.1111/j.1749-6632.1989.tb50513.x.
\bibitem{Per:02} G.~Perelman. \newblock The entropy formula for the ricci flow and its geometric   applications. \newblock arXiv:math.DG/0211159, 2002.
\bibitem{Per:03.2} G.~Perelman. \newblock Finite extinction time for the solutions to the ricci flow on certain   three-manifods. \newblock arXiv:math.DG/0307245, 2003.
\bibitem{Per:03.1} G.~Perelman. \newblock Ricci flow with surgery on three-manifolds. \newblock arXiv:math.DG/0303109, 2003.
\bibitem{Qui:82} F.~Quinn. \newblock Ends of {M}aps {III}: dimensions 4 and 5. \newblock {\em J. Diff. Geom.}, {\bf 17}:503 -- 521, 1982.
\bibitem{Rol:76} D.~Rolfson. \newblock {\em Knots and Links}. \newblock Publish or Prish, Berkeley, 1976.
\bibitem{Ros:94} J.~Rosenberg. \newblock {\em Algebraic {K}-theory and its application}. \newblock Springer, 1994.
\bibitem{WMAPcompactSpace2008} B.F. Roukema, Z.~Bulinski, A.~Szaniewska, and N.E. Gaudin. \newblock The optimal phase of the generalised poincare dodecahedral space   hypothesis implied by the spatial cross-correlation function of the {WMAP}   sky maps. \newblock {\em Astron. Astrophysics}, {\bf 486}:55--74, 2008. \newblock arXiv:0801.0006 [astro-ph].
\bibitem{Scott1983} P.~Scott. \newblock The geometries of 3-manifolds. \newblock {\em Bull. London Math. Soc.}, {\bf 15}:401--487, 1983.
\bibitem{Tamura1992} I.~Tamura. \newblock {\em Topology of Foliations: An Introduction}. \newblock Translations of Math. Monographs Vol. 97. AMS, Providence, 1992.
\bibitem{Thu:72} W.~Thurston. \newblock Noncobordant foliations of {$S^3$}. \newblock {\em BAMS}, 78:511 -- 514, 1972.
\bibitem{Thu:97} W.~Thurston. \newblock {\em Three-Dimensional Geometry and Topology}. \newblock Princeton University Press, Princeton, first edition, 1997.
\bibitem{Whitehead35} J.~H.~C. Whitehead. \newblock A certain open manifold whose group is unity. \newblock {\em Quart. J. Math. Oxford}, 6:268--279, 1935.
\bibitem{Wit:82a} E.~Witten. \newblock Supersymmetry and morse theory. \newblock {\em J. Diff. Geom.}, {\bf 17}:661--692, 1982.
\bibitem{Wit:89.2} E.~Witten. \newblock 2+1 dimensional gravity as an exactly soluble system. \newblock {\em Nucl. Phys.}, {\bf B311}:46--78, 1988/89.
\bibitem{Wit:89.3} E.~Witten. \newblock Topology-changing amplitudes in 2+1 dimensional gravity. \newblock {\em Nucl. Phys.}, {\bf B323}:113--140, 1989.
\bibitem{Wit:91.2} E.~Witten. \newblock Quantization of {C}hern-{S}imons gauge theory with complex gauge   group. \newblock {\em Comm. Math. Phys.}, {\bf 137}:29--66, 1991.
\bibitem{York1972} J.W. York. \newblock Role of conformal three-geometry in the dynamics of gravitation. \newblock {\em Phys. Rev. Lett.}, 28:1082--1085, 1972.
\end{thebibliography}
\end{document}